\documentclass[letterpaper,journal,twoside]{IEEEtran}
\usepackage{cite}
\usepackage{amsmath,amssymb,amsfonts}
\usepackage{algorithmic}
\usepackage{graphicx}
\graphicspath{{figs}{auth}{_aux}}
\usepackage{textcomp}
\def\BibTeX{{\rm B\kern-.05em{\sc i\kern-.025em b}\kern-.08em
		T\kern-.1667em\lower.7ex\hbox{E}\kern-.125emX}}
\usepackage{newtxmath} % math font
\usepackage{bm} % greek bold math font
\usepackage{arydshln}
\usepackage{wasysym}
\usepackage[OT2,OT1]{fontenc}
\newcommand\cyr
{\renewcommand\rmdefault{wncyr}
	\renewcommand\sfdefault{wncyss}
	\renewcommand\encodingdefault{OT2}
	\normalfont
	\selectfont}
\DeclareTextFontCommand{\textcyr}{\cyr}

\newcommand{\SIGMA}{\sigma}
\renewcommand{\SIGMA}{\omega}
\newcommand{\SIGMAR}{\sigma_{4}}
\renewcommand{\SIGMAR}{\sigma}
\newcommand{\SIGMARS}{\sigma_{4*}}
\renewcommand{\SIGMARS}{\sigma_{*}}

\newcommand{\DSIGMAR}{\dot{\sigma}_{4}}
\renewcommand{\DSIGMAR}{\dot{\sigma}}
\usepackage[dvipsnames]{xcolor}
\usepackage{lipsum}

\usepackage[colorlinks=true,allcolors=blue]{hyperref}
\begin{document}
\title{Steering Control of an Autonomous Unicycle}
\author{M\'at\'e B. Vizi, G\'abor Orosz, D\'enes Tak\'acs and G\'abor St\'ep\'an
	\thanks{The research reported in this paper has been supported by the Hungarian National Science Foundation under Grant No. NKFI K 132477, NKFI KKP 133846 and by the HUN-REN Hungarian Research Network; and also by Project no.~TKP-6-6/PALY-2021 provided by the Ministry of Culture and Innovation of Hungary from the National Research, Development and Innovation Fund, financed under the TKP2021-NVA funding scheme. The research of D. T. was supported by a J\'anos Bolyai Research Scholarship of the Hungarian Academy of Sciences.
	}
	\thanks{M\'at\'e B. Vizi and D\'enes Tak\'acs are with the Department of Applied Mechanics, Budapest University of Technology and Economics, Budapest, Hungary and HUN-REN--BME Dynamics of Machines Research Group, Budapest, Hungary (e-mails: vizi@mm.bme.hu, takacs@mm.bme.hu)}
	\thanks{G\'abor Orosz is with the Department of Mechanical Engineering and with the
		Department of Civil and Environmental
		Engineering, University of Michigan, Ann
		Arbor, MI 48109, USA (e-mail: orosz@umich.edu).}
	\thanks{G\'abor St\'ep\'an is with the Department of Applied Mechanics, Budapest University of Technology and Economics, Budapest, Hungary (e-mail: stepan@mm.bme.hu).}
}

\maketitle

\begin{abstract}
	The steering control of an autonomous unicycle is considered. 
	The underlying dynamical model of a single rolling wheel is discussed regarding the steady state motions and their stability. 
	The unicycle model is introduced as the simplest possible extension of the rolling wheel where the location of the center of gravity is controlled.
	With the help of the Appellian approach, a state space representation of the controlled nonholonomic system is built in a way that the most compact nonlinear equations of motions are constructed.
	Based on controllability analysis, feedback controllers are designed which successfully carry out lane changing and turning maneuvers. 
	The behavior of the closed-loop system is demonstrated by numerical simulations.
\end{abstract}


\section{Introduction}\label{sec:introduction}

\begin{figure}[t]
	\centering
	\includegraphics[scale=0.99999]{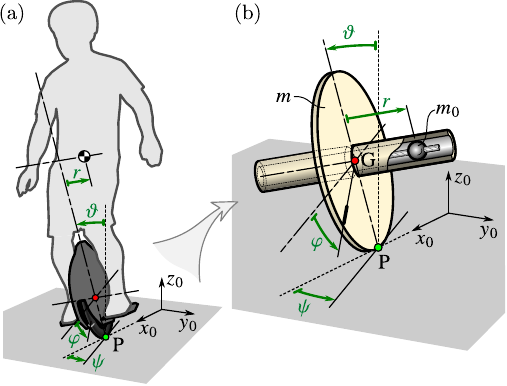}
	\caption{Human riding a unicycle (a), and a simple autonomous EUC (b).}
	\label{fig:howtomodel}   
	\vspace{-4mm}
\end{figure}

\IEEEPARstart{M}{icro-mobility} solutions are spreading rapidly in urban environments \cite{dozza2023micromobility}. 
Among these, human-ridden electric unicycles (EUCs) become more and more popular transportation devices; see Figure~\ref{fig:howtomodel}(a).
These micro-mobility vehicles can match the speed of automobiles in urban traffic while their compact size make them appealing for commute in congested environments.
Due to the three dimensional spatial rolling of the wheel and the stabilization of an unstable equilibrium, the unique dynamics of the unicycle combines agility and maneuverability. 
To exploit these properties, one may consider making EUCs autonomous (see Figure~\ref{fig:howtomodel}(b)) which opens up a challenging avenue for modeling, dynamics and control.

During the last few decades, several autonomous unicycle designs have appeared in the literature which differ on various aspects such as the number and/or types of actuators that can be used for control.
The first publication related to autonomous unicycles known to the authors is \cite{Ozaka_1980_stability} in which the longitudinal/pitch motion is controlled by balancing an inverted pendulum, and the lateral/tilt motion is controlled by moving a mass perpendicular to the wheel.
Two other approaches are presented in \cite{Schoonwinkel_1987}.
In the first case, the longitudinal/pitch motion is also controlled by balancing an inverted pendulum, while the turning/yaw motion of the unicycle is controlled by an overhead flywheel. 
In the second case, the tilt is controlled by adding a second pendulum swinging in the lateral plane.
The overhead flywheel approach was further explored in \cite{Vos_VonFlotow_1990,Naveh_BarYoseph_Halevi_1999}; the lateral pendulum approach can be found in  \cite{Zenkov_Bloch_Marsden_2002}.
The tilt motion of the unicycle can also be controlled by a lateral flywheel, see, for example, \cite{Isomi_Majima_2009,Ruan_2009_Modeling,Han_2014_Balancing},
while the combination of overhead and lateral flywheels can be found in \cite{Zhao_2015_Dynamics,Orosz2023_2}.
Furthermore, the application of gyroscopes for lateral stabilization and steering is presented in \cite{Brown_1996_Single,Dao_2005_Gain,Ghaffari_2010_Improving,Jin_2016_Unicycle}.
Humanoid-type autonomous unicycles are introduced and analyzed in \cite{Sheng_Yamafuji_1995,Suzuki_Moromugi_Okura_2014}, see Figure~\ref{fig:howtomodel}(a) as illustration.

{
	Most of the above approaches provide satisfactory dynamic behavior, but their complexity prohibits closed-form analysis of the controlled system. 
	Our goal here is to develop a simple autonomous unicycle model that is capable of carrying out a variety of maneuvers while it can still be investigated analytically.
	Thus, we consider the simple mechanical model shown in Figure~\ref{fig:howtomodel}(b), which consists of a rolling wheel and added mass that can be moved along the axle to balance the lateral motion.
	We derive the equations of motion for this simple system, analyze the stability properties, and design controllers for maneuvering the autonomous unicycle.
}

The rolling of the wheel can be described using kinematic constraints  \cite{MeiPapRuiSch_2007,Dankowicz_2005_MAMBO,xiong2020bicycle,Qin_2022}.
Thus, the unicycle is considered as a nonholonomic mechanical system. 
Such systems are often described by the generalized Lagrangian equations of the second kind (or Routh--Voss equations) \cite{Routh_1884,Voss_1885}.
This method yields a differential-algebraic system of equations. 
However, eliminating the algebraic variables to obtain a system of ordinary differential equations that is appropriate for control design is a challenging task.
Alternatively, for conservative nonholonomic systems, a Routhian-like model reduction technique, the so-called Lagrange--d'Alembert--Poincar\'{e} equations   can be used to identify conserved quantities and also to explore  symmetries \cite{bloch1996nonholonomic}.

The Appellian approach \cite{Appell_1900,Gibbs_1879,Qin_2022}, which is used in this study, results in a system of first order ordinary differential equations as a compact and simple representation of the underlying nonholonomic system.
Moreover, an innovative definition of the pseudovelocities can significantly reduce the algebraic complexity of the resulting equations of motion while describing the same dynamical system, which simplifies the subsequent analysis.
This enables one to deploy a plethora of control techniques. 
The cost of the Appellian formalism is that accelerations must be calculated, however, the mentioned benefits justify this cost in the examples considered here.

Further details about nonholonomic systems may be found in \cite{Voronets1901,Hamel1938,Kane1961,Gantmacher_1970,NeiFuf1972,Koon_Marsden_1997,OstAng1998,Bloch_2003}.

In this study, we first omit the mass moving along the axle in Figure~\ref{fig:howtomodel}(b) and explore the dynamics of the uncontrolled rolling wheel. 
We categorize different steady state motions (e.g., straight rolling, turning) of interest. 
A strong sufficient condition of stable rolling was given by Bloch in~\cite{Bloch_2003},
while here the local necessary and sufficient condition is derived.
This simplified case also enables us to explain the self stabilizing effects in the tilt direction above a critical speed.
By adding the moving mass and an internal force between the wheel and the mass, 
we create a control system and study how the steady states are affected.
This enables us to design feedback controllers which stabilize the steady states at any speed and enables the unicycle to perform maneuvers such as lane changes and sharp turns.
Our control design exploits the inherent instabilities of the system in order to demonstrate high level of maneuverability.

The article is structured as follows. 
In Section~\ref{sec:prelim}, the modeling framework and steady state analysis of the rolling wheel example are presented. 
Section~\ref{sec:unic_model} introduces a novel autonomous unicycle model and analyzes the steady states of the open-loop system.
Section~\ref{sec:unic_control} proposes controller designs that successfully perform lane changing and turning maneuvers.
The performance of these controllers are demonstrated by numerical simulations. 
We conclude our results in Section~\ref{sec:conclusion} and provide future research directions.

\section{Dynamics of the rolling wheel}\label{sec:prelim}

We introduce the modeling framework and notation on the  rolling wheel example which represents the uncontrolled behavior of the unicycle.
We reveal the dynamical characteristics, including the self-stabilization phenomenon, which can be exploited  for control design.

\subsection{Governing equations}

\begin{figure}[t]
	\centering \includegraphics[width=0.95\columnwidth]{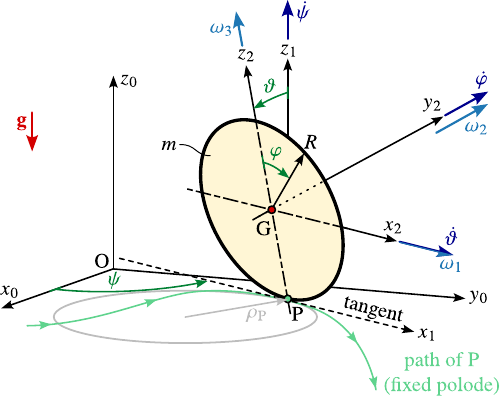}
	\caption{Mechanical model of the rolling wheel}
	\label{fig:disc_mech_model}
	\vspace{-4mm}
\end{figure}

A wheel has ${N=6}$ degrees of freedom (DoF) in the three dimensional space. 
That is, the spatial position and orientation are described by six variables: the position of the center of gravity ${\mathbf r_{\rm G} = [x_{\rm G}\,\ y_{\rm G}\,\ z_{\rm G}]^\mathsf T}$ and the yaw ($\psi$), tilt ($\vartheta$) and pitch ($\varphi$) angles; see Figure~\ref{fig:disc_mech_model}.
Note that tilt is often called roll in the vehicle dynamics literature but here we do not use this convention to avoid confusion with the fact that disc rolls on the horizontal plane.

To describe the motion of the wheel, three coordinate frames are introduced; see Figure~\ref{fig:disc_mech_model}.
The axes $x_0$ and $y_0$ of the ground fixed frame ${\mathrm F_0}$ span the horizontal plane and its $z_0$ axis gives the vertical direction.
The frame ${\mathrm F_1}$ is moving with the wheel such that its origin is the wheel-ground contact point P. 
This frame is rotated with respect to ${\mathrm F_0}$ around the $z_0$ axis with yaw angle $\psi$ so that the $x_1$ axis is tangential to the path of P (the fixed polode).
The frame ${\mathrm F_2}$ is rotated  with respect to ${\mathrm F_1}$ around the $x_1$ axis with the tilt angle $\vartheta$ so the $x_1$ and $z_1$ axes span the plane of the wheel and the $y_2$ axis is aligned with the wheel axle. 
The origin of frame ${\mathrm F_2}$ is placed at the wheel center point $\mathrm G$.

Assume that the wheel rolls without slipping; the kinematic condition of rolling is that the instantaneous center of rotation coincides with the contact point $\rm P$:
\begin{equation}\label{eq:vP}
	\mathbf v_{\rm P} = \mathbf 0\,.%
\end{equation}
This yields the kinematic constraints:%
\begin{align}
	\label{eq:kin_constr}
	\begin{split}
		\dot x_{\rm G} &= \dot \psi R \cos{\psi} \sin{\vartheta} + \dot \vartheta R \sin{\psi} \cos{\vartheta} +  \dot \varphi R \cos{\psi}\,, 
		\\
		\dot y_{\rm G} &= \dot \psi R \sin{\psi} \sin{\vartheta}  - \dot \vartheta R \cos{\psi} \cos{\vartheta} + \dot \varphi R \sin{\psi}\,, 
	\end{split}
\end{align}
and the geometric constraint
\begin{align}
	\dot z_{\rm G} &= - R  \dot \vartheta \sin{\vartheta}\, \quad\Rightarrow\quad  z_{\rm G} = R \cos{\vartheta}\,,
	\label{eq:geom_constr_eq_z}
\end{align}
where the vertical position $z_{\rm G}$ depends on the tilt angle $\vartheta$; see Figure~\ref{fig:disc_mech_model}. Therefore the rolling wheel is a nonholonomic mechanical system with ${n_{\rm g}=1}$ geometric constraint \eqref{eq:geom_constr_eq_z} and ${n_{\rm k}=2}$ kinematic constraints \eqref{eq:kin_constr}.
Note that the geometric and kinematic constraints are also referred to as holonomic and nonholonomic constraints, respectively.
The equations of motion are derived in Appendix~\ref{app:rollingdisc_model_deriv}
using the Appellian approach \cite{Appell_1900, Gibbs_1879, Qin_2022} to provide the most compact algebraic form.

According to the number of geometric and kinematic constraints, ${n_q=6-n_{\rm g}=5}$ generalized coordinates
have to be chosen to describe the system unambiguously; let these be:
\begin{equation}
	\big( x_{\rm G}, \, y_{\rm G}, \, \psi, \, \vartheta, \, \varphi \big)\,.
\end{equation}
Moreover, ${n_\sigma = n_q - n_{\rm k} = 3}$ pseudovelocities 
have to be chosen; let these be defined by the components of the angular velocity  $\bm\omega$ resolved in frame ${\rm F}_{2}$ (cf.~\eqref{eq:omega} in Appendix~\ref{app:rollingdisc_model_deriv}):
\begin{equation}\label{eq:sigmadef_disc}
	\SIGMA_1 := \dot\vartheta\,\!,\quad 
	\SIGMA_2 := \dot\psi\sin\vartheta + \dot\varphi\,,\quad
	\SIGMA_3 := \dot\psi\cos\vartheta\,\!. 
\end{equation}
Then, the Appellian approach yields the equations of motion:
\begin{equation}
	\label{eq:eq_mot_disc}
	\begin{split}
		&\begin{cases}\begin{aligned}
				\dot{\SIGMA}_1 &= \frac{6}{5} \SIGMA_2 \SIGMA_3 - \frac{1}{5} \SIGMA_{3}^{2} \tan \vartheta + \frac{4 g}{5 R} \sin\vartheta \,, %
				\\
				\dot{\SIGMA}_2 &= - \frac{2}{3} \SIGMA_1 \SIGMA_3\,,
				\\
				\hspace{0.1ex}\dot{\SIGMA}_3 &= - 2 \SIGMA_1 \SIGMA_2 + \SIGMA_1 \SIGMA_3 \tan\vartheta\,\!,    
				\\
				\dot{\vartheta} &= \SIGMA_1\,, %
			\end{aligned}
		\end{cases}
		\\
		&\begin{cases}
			\begin{aligned}
				\dot{\psi} &= \SIGMA_3\frac{1}{\cos\vartheta}\,, %
				\\
				\dot{\varphi} &= \SIGMA_2 - \SIGMA_3 \tan\vartheta\,\!,
				\\
				\dot{x}_{\rm G} &= \SIGMA_1  R \sin\psi \cos\vartheta + \SIGMA_2 R \cos\psi\,,
				\\
				\dot{y}_{\rm G} &= - \SIGMA_1 R \cos\psi \cos\vartheta + \SIGMA_2 R \sin\psi\,,
			\end{aligned}
		\end{cases}
	\end{split}
\end{equation}
that is, the rolling wheel is an ${n =6 - n_{\rm g} -n_{\rm k}/2 = 4}$ DoF nonholonomic mechanical system. 
The equations in \eqref{eq:eq_mot_disc} are ordered such that the system can be separated into essential dynamics (the first four equations) and hidden dynamics (the second four equations) where the essential dynamics is independent of the hidden dynamics \cite{Routh_1884}.
The equations are in the form ${\dot{\mathbf x} = f(\mathbf{x})}$ where the state is defined as
\begin{equation} \label{eq:wheel_state_vector}
	\mathbf x
	=
	\left[\begin{array}{cccc;{2pt/2pt}cccc}
		\SIGMA_{1}
		&\SIGMA_{2}
		&\SIGMA_{3}
		&\vartheta
		&\psi
		&\varphi
		&x_{\rm G}
		&y_{\rm G}
	\end{array}\right]^{\mathsf T}\!,
\end{equation}
{%
	where the dashed line separates the essential states from the cyclic coordinates describing hidden motion.%
}%

\subsection{Steady state motions}\label{subsec:teady_state_disc}

The rolling wheel exhibits a steady state motion when the essential dynamics (first four equations in \eqref{eq:eq_mot_disc}) possess an equilibrium.
That is, the pseudovelocities
and the tilt angle are constants:
\begin{equation}
	\label{eq:essentialdyn_fixpoint_disc}
	\SIGMA_1(t)      \equiv   \SIGMA_{1*}\,,\ 
	\SIGMA_2(t)       \equiv   \SIGMA_{2*}\,,\ 
	\SIGMA_3(t)       \equiv   \SIGMA_{3*}\,,\ 
	\vartheta(t)      \equiv  \vartheta_{*}\,.
\end{equation}
Then, according to the fourth equation in \eqref{eq:eq_mot_disc}, the tilt rate must be zero, i.e., ${\SIGMA_{1*} = 0}$. Substituting this into \eqref{eq:eq_mot_disc} the first equation yields
\begin{equation}
	\label{eq:steadystaterelation_pseudoveloc_disc}
	\frac{6}{5} \SIGMA_{2*} \SIGMA_{3*}  - \frac{1}{5} \SIGMA_{3*}^2 \tan{\vartheta_{*} } + \frac{4 g}{5 R} \sin{ \vartheta_{*} } = 0\,,
\end{equation}
while the hidden motion can be expressed as
\begin{equation}
	\label{eq:const_gen_velocs_disc}
	\arraycolsep=1pt
	\begin{array}{rlrl}
		\dot \psi(t)       &\equiv \dot{\psi}_{*}        =   \SIGMA_{3*}\frac{1}{\cos{\vartheta_{*}}}\,, \ 
		&
		\dot x_{\rm G}(t)  &=  \SIGMA_{2*} R \cos{\psi}(t)\,, 
		\\
		\dot \varphi(t)    &\equiv \dot{\varphi}_{*}     =   \SIGMA_{2*} -  \SIGMA_{3*} \tan{\vartheta_{*}}\,, \ \ 
		&
		\dot y_{\rm G}(t)  &=  \SIGMA_{2*} R \sin{\psi}(t) \,.
	\end{array}
\end{equation}
That is, the yaw rate $\dot \psi$ and the pitch rate $\dot \varphi$ are constants,
while the horizontal velocity components $\dot x_{\rm G}$, $\dot y_{\rm G}$ of the center of gravity vary with time through the yaw angle $\psi$.

Integrating \eqref{eq:const_gen_velocs_disc}, the generalized coordinates become
\begin{align}
	\label{eq:steadystate_hiddendyn_disc}
	\begin{split}
		\psi_{*}(t)       &= \dot{\psi}_{*} t + \psi_0   \,, 
		\\
		\varphi_{*}(t)    &= \dot{\varphi}_{*} t + \varphi_0\,, 
		\\
		x_{\rm G*}(t)  &= \begin{cases}
			\bigg(\dfrac{\dot{\varphi}_{*}}{\dot{\psi}_{*}} + \sin\vartheta_{*}\bigg)
			R \sin{(\dot{\psi}_{*} t + \psi_0)} + x_0 & {\rm if }\ \ \dot{\psi}_{*} \neq 0\,,
			\\
			\dot{\varphi}_{*} t R \cos\psi_0 + x_0 & {\rm if }\ \ \dot{\psi}_{*} = 0\,,
		\end{cases}
		\\
		y_{\rm G*}(t)  &= \begin{cases}
			-\bigg(\dfrac{\dot{\varphi}_{*}}{\dot{\psi}_{*}} + \sin\vartheta_{*}\bigg)
			R \cos{(\dot{\psi}_{*} t + \psi_0)} + y_0 & {\rm if }\ \ \dot{\psi}_{*} \neq 0\,,
			\\
			\dot{\varphi}_{*}  t R \sin\psi_0 + y_0 & {\rm if }\ \ \dot{\psi}_{*} = 0\,,
		\end{cases} %
	\end{split}
\end{align}
where the  $\psi_0$ and $\varphi_0$ denote the initial yaw and pitch angles while $x_0$ and $y_0$ originate in the initial position of point $\rm G$.
The center of gravity $\rm G$  follows a circular path of radius ${\rho_{\rm G} = |\dot{\varphi}_{*}/\dot{\psi}_{*} + \sin\vartheta_{*}|\,R }$
if the yaw rate is not zero (${\dot{\psi}_{*}\neq 0}$). Correspondingly, the contact point P draws a circle of radius 
${
	\rho_{\rm P} = \left| {\dot{\varphi}_{*}}/{\dot{\psi}_{*}} \right| R
}$
on the ground plane; see Figure~\ref{fig:disc_mech_model}. 
We refer to this motion as turning-rolling in the rest of the paper.
For zero yaw rate (${\dot{\psi}_{*} = 0}$), the center or gravity moves along a straight path.
We refer to this as straight rolling in the rest of the paper.

Keep in mind that the steady state tilt angle, yaw rate and pitch rate are not independent of each other.
To establish a relationship between these quantities, one must replace the steady state pseudovelocities 
($\SIGMA_{1*}$, $\SIGMA_{2*}$, $\SIGMA_{3*}$)
in~\eqref{eq:steadystaterelation_pseudoveloc_disc} with the generalized velocities
($\dot\psi_{*}$, $\dot\varphi_{*}$)
according to \eqref{eq:sigmadef_disc}. 
This yields
\begin{equation}\label{eq:steadystaterelation_genveloc_disc}
	\dot\psi_{*}^{2} \sin{\vartheta_{*}} \cos{\vartheta_{*}} 
	+ \frac{6}{5} \dot{\psi}_{*} \dot{\varphi}_{*} \cos{\vartheta_{*}} 
	+ \frac{4 g}{5 R} \sin{\vartheta_{*}} = 0\,, 
\end{equation}
which is visualized in Figure~\ref{fig:levels_and_stability_disc}(a).
Since \eqref{eq:steadystaterelation_genveloc_disc} is linear in the pitch rate $\dot{\varphi}_{*}$, one may express $\dot{\varphi}_{*}$ as a function of the the tilt angle $\vartheta_{*}$ and the yaw rate $\dot{\psi}_{*}$ for turning-rolling (${\dot{\psi}_{*}\neq 0}$).

For straight rolling (${\dot{\psi}_{*}=0}$), \eqref{eq:steadystaterelation_genveloc_disc} 
reduces to 
\begin{equation}
	\label{eq:straightrolling_disc}
	\sin{\vartheta_{*}} = 0,
\end{equation}
yielding ${\vartheta_{*}=0}$, that is, the wheel must be non-tilted.
Observe that \eqref{eq:straightrolling_disc} is independent of the pitch rate $\dot{\varphi}_{*}$, so the straight rolling is feasible for arbitrary pitch rates.
Another special case is when the pitch rate is zero (${\dot{\varphi}_{*}=0}$).
Then~\eqref{eq:steadystaterelation_genveloc_disc} results in 
\begin{equation}
	\label{eq:steadystate_spinning_disc}
	\dot\psi_{*}^{2} \sin{\vartheta_{*}} \cos{\vartheta_{*}} 
	+ \dfrac{4 g}{5 R} \sin{\vartheta_{*}} = 0\,,
\end{equation}
which can only hold when the tilt angle is zero, that is, ${\vartheta_{*}=0}$.
We refer to this solution as spinning on the spot in the rest of the paper.
Finally, the very special case ${\dot{\psi}_{*} = \dot{\varphi}_{*} = 0}$ corresponds to the  static equilibrium of the standing disc, which is, indeed, unstable.

The stability of the steady states (turning-rolling, straight rolling and spinning) are analyzed consecutively below.

\subsection{Stability of the steady state motions}

The steady state motions of the rolling wheel are defined with respect to the essential motion independently of the cyclic states of the system. 
Accordingly, when we talk about the stability of the steady state motions, it is only about the essential dynamics, while the hidden motions may present instability in the Lyapunov sense.

Let us consider two examples when the wheel initially rolls straight along the $x$ axis.
In the first case, a small perturbation may increase the speed of rolling. 
If the essential states ${\SIGMA_{1}, \SIGMA_{2}, \SIGMA_{3} ,\vartheta}$ remain in the small vicinity of the original values then the essential motion is stable.
However, the increased velocity $\omega_{2}$ causes the cyclic coordinates $\varphi, x_{\mathrm G}$ move away linearly in time from the original unperturbed trajectories; therefore, the hidden motion is unstable.
In the second case, let the perturbation cause a small tilt angle ${\vartheta_0}$, tilt rate ${\dot\vartheta_0=\SIGMA_{10}}$ and/or a small yaw rate ${\dot\psi_0 = \SIGMA_{30} }$. 
If the essential states remain in the vicinity of the original unperturbed values then the essential motion is stable.
However, this small perturbation yields a circular motion of the wheel with large radius instead of the straight motion, so the hidden motion is unstable.

\color{black}

For simplicity assume the initial values ${\psi_0 = 0}$, ${\varphi_0 = 0}$, ${x_0 = 0}$, and ${y_0 = 0}$.
According to \eqref{eq:sigmadef_disc}, \eqref{eq:essentialdyn_fixpoint_disc}, and \eqref{eq:steadystate_hiddendyn_disc}, the steady state motion is given as
\begin{equation}
	\label{eq:state_vec_disc}
	{\mathbf x}_{*}  = 
	\left[\begin{array}{c}
		\SIGMA_{1*}\\
		\SIGMA_{2*}\\
		\SIGMA_{3*}\\
		\vartheta_{*}\\\hdashline[2pt/2pt]
		\psi_{*}(t)\\
		\varphi_{*}(t)\\
		x_{\rm G *}(t)\\
		y_{\rm G *}(t)
	\end{array}\right]
	=\left[\begin{array}{c}
		0
		\\
		\dot{\psi}_{*}\sin\vartheta_{*} + \dot{\varphi}_{*}
		\\
		\dot{\psi}_{*}\cos\vartheta_{*}
		\\
		\vartheta_{*}
		\\\hdashline[2pt/2pt]
		\dot{\psi}_{*} t 
		\\
		\dot\varphi_{*}t
		\\
		\dots %
		\\
		\dots %
	\end{array}\right]\,,
\end{equation}
where the dots in $x_{\rm G *}(t)$ and $y_{\rm G *}(t)$ refer to \eqref{eq:steadystate_hiddendyn_disc}.

Introducing the state perturbations ${\tilde{\mathbf x} := \mathbf x - {\mathbf x}_{*}}$, \eqref{eq:eq_mot_disc} leads to the linearized equations of the form
${\dot{\tilde {\mathbf x}} = \mathbf{A} \tilde{\mathbf x}}$
where the state matrix is obtained as  ${\mathbf A = f'({\mathbf x}_{*})}$. This  yields

\begin{equation}
	\label{eq:statematrixA_disc}
	\mathbf A =  \left[\begin{array}{cccc;{2pt/2pt}cccc}
		0      & A_{12} & A_{13} & A_{14} & 0      & 0 & 0 & 0 \\ %
		A_{21} & 0      & 0      & 0      & 0      & 0 & 0 & 0 \\ %
		A_{31} & 0      & 0      & 0      & 0      & 0 & 0 & 0 \\ %
		1      & 0      & 0      & 0      & 0      & 0 & 0 & 0 \\ \hdashline[2pt/2pt] %
		0      & 0      & A_{53} & A_{54} & 0      & 0 & 0 & 0 \\ %
		0      & 1      & A_{63} & A_{64} & 0      & 0 & 0 & 0 \\ %
		A_{71} & A_{72} & 0      & 0      & A_{75} & 0 & 0 & 0 \\ %
		A_{81} & A_{82} & 0      & 0      & A_{85} & 0 & 0 & 0
	\end{array}\right],
\end{equation}%
where
\begin{equation}\label{eq:statematrixA_disc2}
	\def\arraystretch{1.5}
	\begin{array}{llllll}
		\multicolumn{3}{l}{   A_{12} =\frac{6 }{5}\dot{\psi}_{*} \cos{\vartheta_{*}}\,,} 
		& \multicolumn{3}{l}{ A_{13} = \frac{4}{5} \dot{\psi}_{*} \sin{\vartheta_{*}} + \frac{6}{5} \dot{\varphi}_{*}\,,}
		\\
		\multicolumn{3}{l}{   A_{14} = - \frac{1}{5}\dot\psi_{*}^{2} + \frac{ 4 g }{5 R} \cos{\vartheta_{*}} \,,}             
		& \multicolumn{3}{l}{ A_{21} = - \frac{2}{3} \dot{\psi}_{*} \cos{\vartheta_{*}}}
		\\
		\multicolumn{3}{l}{   A_{31} = - \dot{\psi}_{*} \sin{\vartheta_{*}} - 2 \dot{\varphi}_{*} \,,}     
		& \multicolumn{3}{l}{ A_{53} = \frac{1}{\cos{\vartheta_{*}}} \,,} 
		\\
		\multicolumn{3}{l}{   A_{54} = \dot{\psi}_{*} \tan{\vartheta_{*}} \,,} 
		& \multicolumn{3}{l}{ A_{63} = - \tan{\vartheta_{*}} \,,} 
		\\
		\multicolumn{3}{l}{   A_{64} = - \frac{\dot{\psi}_{*}}{\cos{\vartheta_{*}}} \,,} 
		&\multicolumn{3}{l}{  A_{71} = R \sin\psi_{*}\cos{\vartheta_{*}} \,,}            
		\\
		\multicolumn{2}{l}{   A_{72} = R \cos \psi_{*} \,,   }                     
		& \multicolumn{4}{l}{ A_{75} = - R \sin\psi_{*} \left(\dot{\psi}_{*} \sin{\vartheta_{*}} + \dot{\varphi}_{*}\right)\!,} 
		\\
		\multicolumn{3}{l}{   A_{81} = - R \cos\psi_{*}\cos{\vartheta_{*}}\,,}
		& \multicolumn{2}{l}{ A_{82} = R \sin \psi_{*} \,,   }     
		\\
		\multicolumn{4}{l}{   A_{85} = R \cos\psi_{*} \left(\dot{\psi}_{*} \sin{\vartheta_{*}} + \dot{\varphi}_{*}\right),}
	\end{array}
\end{equation}
and the pitch rate $\dot\varphi _{*}$ must be substituted according to 
\eqref{eq:steadystaterelation_genveloc_disc}.
The dashed lines distinguish the parts of ${ \mathbf A }$ related to the essential and to the hidden dynamics. 

The linear stability of the steady state 
(essential)
motion can be determined by the characteristic equation
${\det(\lambda \mathbf I - \mathbf{\hat{A}}) = 0,}$
where $\mathbf{\hat{A}}$ is the top quadrant of $\mathbf A$ in \eqref{eq:statematrixA_disc}.  
This yields
\begin{equation} \label{eq:wheelChp}
	\big(\lambda^{2} - A_{12} A_{21} - A_{13} A_{31} - A_{14}\big) 
	{\lambda^{2} }
	= 0\,,
\end{equation}
and the characteristic roots become 
\begin{equation}
	\label{eq:eigvals12_disc_general}
	\begin{split}
		\lambda_{1,2} &= 
		\pm\sqrt{ \frac{4g}{5R} \cos{\vartheta_{*}} - 
			\bigg( \dot\psi_{*}^{2} + \frac{14}{5} \dot{\psi}_{*} \dot{\varphi}_{*} \sin{\vartheta_{*}} + \frac{12}{5} \dot\varphi_{*}^{2} \bigg) }\,,    
		\\
		\lambda_{3,4} &= 0. %
	\end{split}
\end{equation}
The corresponding steady state  is unstable if the characteristic roots are real, ${\lambda_{1,2} \in\mathbb R}$ and ${\lambda_1 = -\lambda_2}$.
Otherwise, if the first two characteristic roots constitute a purely imaginary complex conjugate pair ${\lambda_{1,2} \in\mathbb{C}}$ and ${\lambda_1 = \bar\lambda_2}$, a `doubtful Lyapunov' case arises due to the zero eigenvalue of multiplicity two. 
One may show, however, that both the algebraic and geometric multiplicity of the zero eigenvalue are two, and
in Appendix~\ref{app:disc_stability} the Lyapunov stability of the essential dynamics is proven.
Note that the full motion is not stable in the Lyapunov sense due to the zero eigenvalue of algebraic multiplicity four and geometric multiplicity one belonging to the hidden dynamics (see the lower right quadrant of state matrix~\eqref{eq:statematrixA_disc}).
\color{black}

To sum up the analysis of the linear model, the necessary and sufficient condition for the stable rolling of the wheel is
\begin{equation}\label{eq:wheelNeccAndSuffCond}
	\dot\psi_{*}^{2} + \frac{14}{5} \dot{\psi}_{*} \dot{\varphi}_{*} \sin{\vartheta_{*}} + \frac{12}{5} \dot\varphi_{*}^{2} -\frac{4g}{5R} \cos{\vartheta_{*}} > 0,
\end{equation}
where \eqref{eq:steadystaterelation_genveloc_disc} holds.
When varying the tilt angle $\vartheta_{*}$, the yaw rate $\dot\psi_{*}$ and pitch rate $\dot \varphi_{*}$,
one may change the stability of the steady-state motion.
\color{black}

\subsubsection{Turning-rolling}

In case of a general turning-rolling type steady state motion, neither the tilt angle nor the yaw and pitch rates are zero, i.e., ${\vartheta_{*}\neq 0}$, ${\dot\psi_{*}\neq 0}$ and ${\dot \varphi_{*}\neq 0}$. 
Then stability of the steady state changes at the critical yaw rates
\begin{equation}
	\label{eq:psi_crit_general_disc}
	\begin{split}
		\dot{\psi}_{\rm crit,1,2} = \sqrt {\frac{2g}{5 R}}
		\sqrt{\frac{3 - 6 \cos^{2}{\vartheta_{*}} \pm \sqrt{76 \sin^{4}{\vartheta_{*} } - 96 \sin^{2}{\vartheta_{*} } + 9} }{(2 \sin^{2}{\vartheta_{*} } - 3) \cos{\vartheta_{*} }}}
	\end{split}
\end{equation}
for 
\begin{equation}
	|\vartheta_{*}| \leq \mathcal V = \arcsin{\textstyle\left(\sqrt{\frac{12}{19} - \frac{9 \sqrt{5}}{38}} \right) \approx 18.62^\circ }.
\end{equation}
For a given tilt angle ${|\vartheta_{*}| \leq \mathcal V}$, 
if ${|\dot{\psi}_{*}| < \dot{\psi}_{\rm crit,1}}$ or ${\dot{\psi}_{\rm crit,2} < |\dot{\psi}_{*}|}$ then the characteristic roots in \eqref{eq:eigvals12_disc_general} are purely imaginary ${\lambda_{1,2} \in\mathbb C}$ and ${\lambda_1 = \bar\lambda_2}$ and the corresponding motion is  stable.
Otherwise, if ${\dot{\psi}_{\rm crit,1} < |\dot{\psi}_{*}| < \dot{\psi}_{\rm crit,2}}$ then the characteristic roots are real ${\lambda_{1,2} \in\mathbb R}$ and ${\lambda_1 = -\lambda_2}$, and the corresponding motion is unstable. 
Notice that, if the tilt angle is large enough ${|\vartheta_{*}| > \mathcal V}$ then all corresponding steady state motion are stable, since $\mathcal V$ is independent of any physical parameter. 
This is a fundamental physical constant related to the steady state stability of any rolling wheel.

\begin{figure*}
	\centering
	\includegraphics[width=0.95\columnwidth]{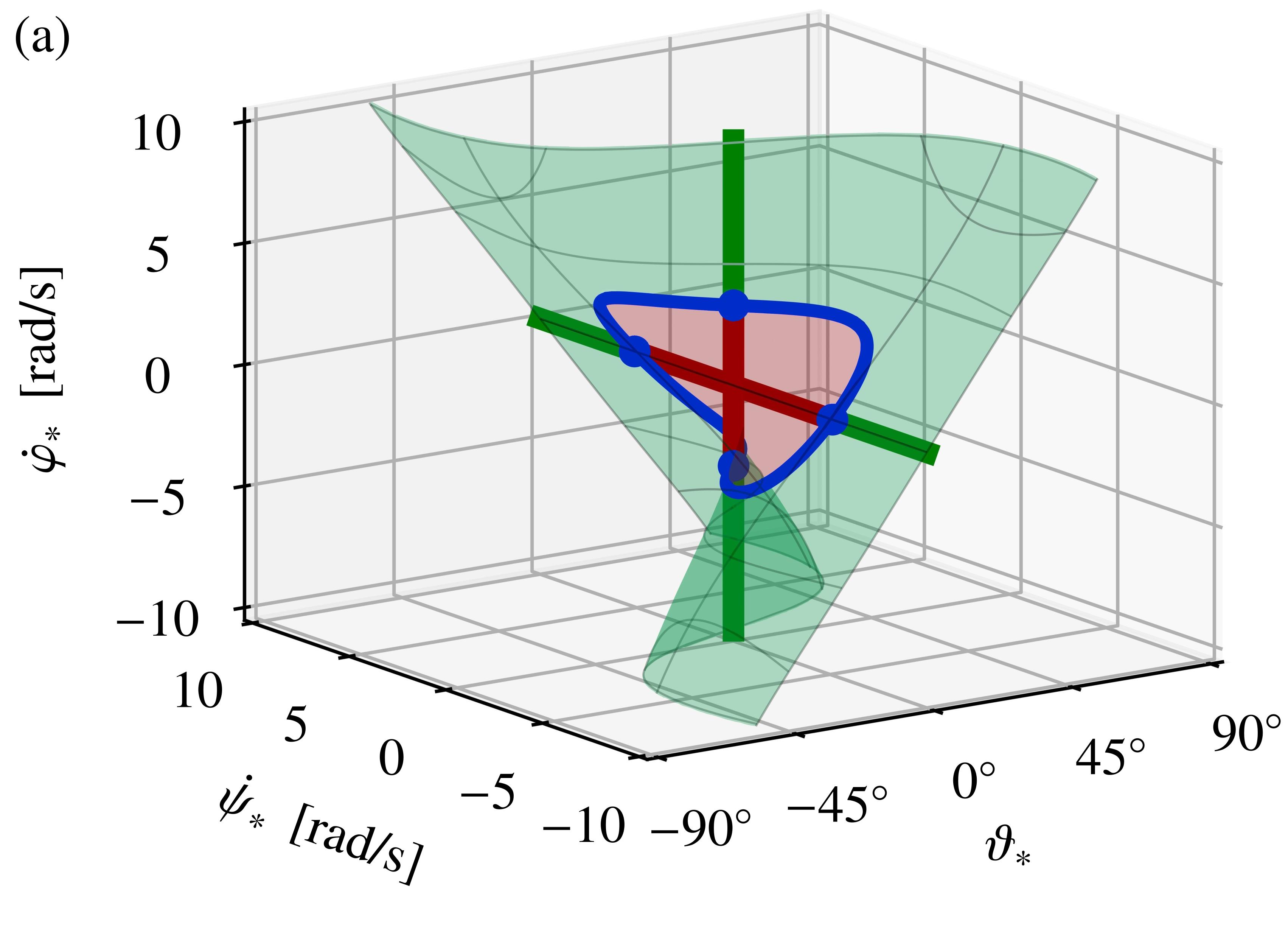}
	\quad
	\includegraphics[width=0.95\columnwidth]{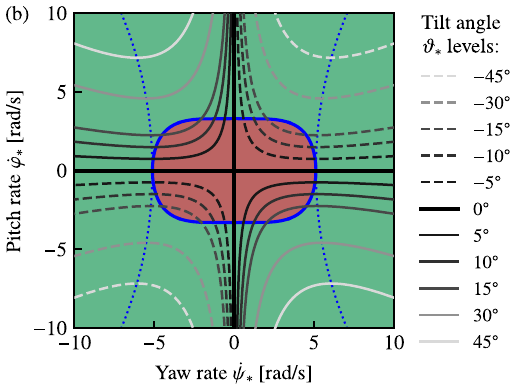}
	\\[1em]
	\includegraphics[width=0.95\columnwidth]{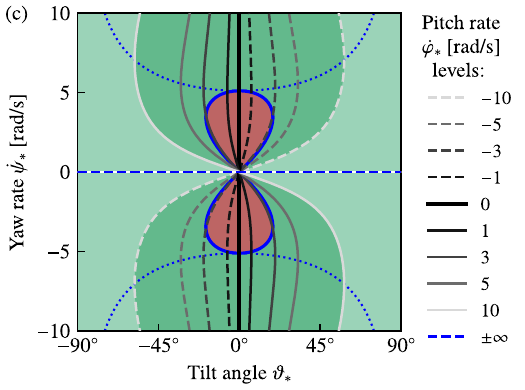}
	\quad
	\includegraphics[width=0.95\columnwidth]{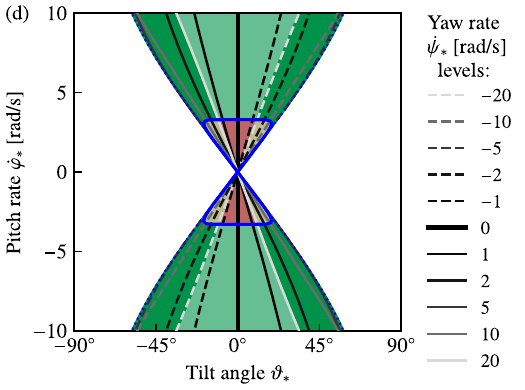}
	\caption{
		Level sets and stability of steady state motion 
		\eqref{eq:steadystaterelation_genveloc_disc} 
		of the rolling wheel: stable and unstable steady states are denoted by green and red, respectively, the blue line is the stability boundary \eqref{eq:wheelNeccAndSuffCond} and the dotted blue line is the folding of the surface. 
		In panel (d), white denotes that no steady state exists, while yellow corresponds to the cases when there are both stable and unstable steady states due to the folding of the surface.
	}
	\label{fig:levels_and_stability_disc}
	\vspace{-4mm}
\end{figure*}

\subsubsection{Straight rolling} 

As discussed above the straight rolling steady state motion (${\dot{\psi}_{*}=0}$) can only occur for zero tilt angle ${\vartheta_{*}=0}$.
That is, the steady state \eqref{eq:state_vec_disc}  simplifies to 
\begin{equation}\label{eq:straight_rolling}
	\SIGMA_{2*} = \dot\varphi_{*}\,, \hspace{1em}
	\varphi_{*}(t) = \dot\varphi_{*}t\,, \hspace{1em}
	x_{\rm G *}(t) = R \dot{\varphi}_{*} t\,, 
\end{equation}
and all the other states become zero.
The state matrix \eqref{eq:statematrixA_disc} is
\begin{equation}
	\label{eq:Amat_straight_disc}
	\mathbf A =  
	\left[\begin{array}{cccc;{2pt/2pt}cccc}%
		0                     & 0 & \frac{6 }{5}\dot{\varphi}_{*} & \frac{4 g}{5 R} & 0                   & 0 & 0 & 0 \\
		0                     & 0 & 0                             & 0               & 0                   & 0 & 0 & 0 \\
		- 2 \dot{\varphi}_{*} & 0 & 0                             & 0               & 0                   & 0 & 0 & 0 \\
		1                     & 0 & 0                             & 0               & 0                   & 0 & 0 & 0 \\\hdashline[2pt/2pt]
		0                     & 0 & 1                             & 0               & 0                   & 0 & 0 & 0 \\
		0                     & 1 & 0                             & 0               & 0                   & 0 & 0 & 0 \\
		0                     & R & 0                             & 0               & 0                   & 0 & 0 & 0 \\
		- R                   & 0 & 0                             & 0               & R \dot{\varphi}_{*} & 0 & 0 & 0
	\end{array}\right]\,,
\end{equation}
and the characteristic roots \eqref{eq:eigvals12_disc_general} simplify to
\begin{equation}
	\lambda_{1,2} = \pm  \sqrt{ \frac{4g}{5 R} - \frac{12}{5}{\dot\varphi}_{*}^{2}}\,.
\end{equation}
This leads to the critical pitch rate
\begin{equation}
	\label{eq:critpitchrate_disc}
	{\dot{\varphi}}_{\rm crit} = \sqrt{\frac{g}{3R}}\,,
\end{equation}
that is, the straight rolling steady state is stable if the wheel rolls fast enough, ${|\dot{\varphi}_{*}|>\dot{\varphi}_{\rm crit}}$.
Otherwise, it is unstable.

\color{black}

We remark that instead of the pitch rate $\dot{\varphi}_{*}$ one may introduce the parameter ${v_{*}= \dot{\varphi}_{*} R}$ which is the steady state velocity of the center of gravity. This way the critical pitch rate \eqref{eq:critpitchrate_disc} can be converted to a critical velocity.

\subsubsection{Spinning on the spot}

The spinning steady state motion (${\dot \varphi_{*} = 0}$) can also occur for zero tilt angle ${\vartheta_{*}=0}$.
Then the steady state becomes
\begin{equation}
	\SIGMA_{3*} = \dot\psi_{*}\,, \hspace{1em} \psi_{*} = \dot\psi_{*}t\,,
\end{equation}
and all the other states are zero. The state matrix \eqref{eq:statematrixA_disc} becomes
\begin{equation}
	\arraycolsep=3.4pt
	\mathbf A =  
	\left[\begin{array}{cccc;{2pt/2pt}cccc}%
		0 & \frac{6}{5} \dot\psi_{*} & 0 & \frac{4 g}{5 R} - \frac{1}{5}\dot\psi_{*}^{2} & 0 & 0 & 0 & 0\\
		- \frac{2}{3}\dot\psi_{*} & 0 & 0 & 0 & 0 & 0 & 0 & 0\\
		0 & 0 & 0 & 0 & 0 & 0 & 0 & 0\\
		1 & 0 & 0 & 0 & 0 & 0 & 0 & 0\\\hdashline[2pt/2pt]
		0 & 0 & 1 & 0 & 0 & 0 & 0 & 0\\0 & 1 & 0 & - \dot\psi_{*} & 0 & 0 & 0 & 0\\
		R\sin\psi_{*} & R\cos\psi_{*} & 0 & 0 & 0 & 0 & 0 & 0\\
		- R\cos\psi_{*} & R\sin\psi_{*} & 0 & 0 & 0 & 0 & 0 & 0   
	\end{array}\right],
\end{equation}
while the characteristic roots \eqref{eq:eigvals12_disc_general} and the critical yaw rate \eqref{eq:psi_crit_general_disc} simplify to
\begin{equation}
	\lambda_{1,2} = \pm\sqrt{\frac{ 4 g}{5 R} - {\dot\psi}_{*}^{2}}\,,
	\qquad
	\dot\psi_{\rm crit} = \sqrt{\frac{4g}{5R}}\,.
\end{equation}
That is, the spinning steady state is stable if the wheel spins fast enough ${|{\dot{\psi}}_{*}| > {\dot{\psi}}_{\rm crit}}$. Otherwise, it is unstable.

\begin{table}[!t]
	\caption{Numerical parameters for the rolling wheel and unicycle}
	\label{tab:physical_consts_disc}
	\setlength{\tabcolsep}{3pt}
	\centering
	\begin{tabular}{llcrl}
		\hline
		\hspace{1ex}\hspace{-4pt} & Quantity                   &    Symbol     & Value & Unit\hspace*{1ex}      \\ \hline
		& wheel mass                 &      $m$      &    10 & kg                     \\
		& point mass                 &    \ $m_0$    &     5 & kg                     \\
		& wheel radius               &      $R$      &   0.3 & m                      \\
		& gravitational acceleration &      $g$      &  9.81 & m/s\textsuperscript{2} \\ \hline
	\end{tabular}
	\vspace{-4mm}
\end{table}

All results about the steady state motions and their stability are shown in Figure~\ref{fig:levels_and_stability_disc}.
Panel (a) shows the steady state motions \eqref{eq:steadystaterelation_genveloc_disc} as surface in the ($\vartheta_{*}$, $\dot{\psi}_{*}$, $\dot{\varphi}_{*}$) space.
Panel (b) shows the level sets of $\vartheta_{*}$ in the ($\dot{\psi}_{*}$, $\dot{\varphi}_{*}$) plane. 
Similarly, panels (c) and (d) show the level sets of $\dot{\psi}_{*}$ and $\dot{\varphi}_{*}$, respectively.
Green (\textcolor[RGB]{102,191,146}{$\blacksquare\!\blacksquare$}) and red (\textcolor[RGB]{191,102,102}{$\blacksquare\!\blacksquare$}) areas represent stable and unstable steady state solutions, respectively, while the solid blue line~(\textbf{\textcolor[RGB]{0,0,255}{---}}) is the stability boundary \eqref{eq:wheelNeccAndSuffCond}.
The dashed blue line~(\textbf{\textcolor[RGB]{0,0,255}{--\,--}}) in panel~(c) shows where ${\dot\varphi_{*} \rightarrow \pm\infty}$.
The dotted blue curves~(\textbf{\textcolor[RGB]{0,0,255}{$\cdots$}}\!) in panels~(b) and (c) indicate the folding of the surface and these bound the hour glass-shaped area in panel (d). 
In the dark green area (\textcolor[RGB]{41,165,104}{$\blacksquare\!\blacksquare$}) in panel (d), two stable steady state states exist with different yaw rates $\dot\psi_{*}$, while in the yellow area (\textcolor[RGB]{189,189,145}{$\blacksquare\!\blacksquare$}) a stable and an unstable steady state exist.
In the white area, no steady state exists regardless of the yaw rate due to the folding of the surface.

\begin{figure}[b]
	\centering
	\includegraphics[width=0.99\columnwidth]{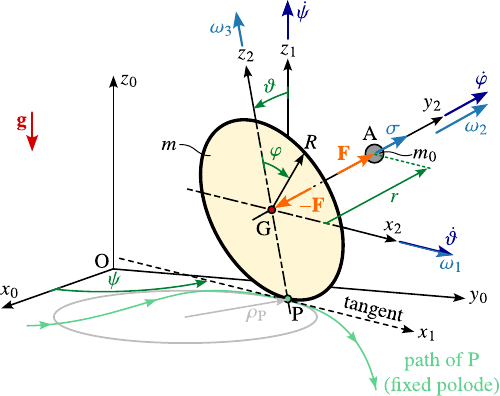}
	\caption{Model of the unicycle for maneuvering}
	\label{fig:unicycle_mech_model}
\end{figure}

\section{Modeling the unicycle}\label{sec:unic_model}

The unicycle is shown in Figure~\ref{fig:unicycle_mech_model}, where the motion is controlled by moving a mass point along the axle of the wheel. In this section, we present the governing equations and analyze the steady states of the open loop system.

\subsection{Governing equations}

The model of the unicycle is built upon the rolling wheel model presented in Section~\ref{sec:prelim} with the added point mass $m_0$ moving along the axle. 
The mass' relative position is given by
\begin{equation} \label{eq:geom_constr_mr_perpendicular}
	\mathbf r_{\rm GA} = \begin{bmatrix}
		0&r&0
	\end{bmatrix}_{{\rm F}_{2}}^{\mathsf T},
\end{equation}
where $r$ represents the position of the point mass along the axle while other two coordinates are constrained at zero by the axle, resulting in two geometric constraints. 
Between the disc and the point mass, there is an internal pair of forces; the control force acting on the point mass is
\begin{equation}
	\label{eq:Fr}
	\mathbf F = \begin{bmatrix}
		0&u&0
	\end{bmatrix}_{{\rm F}_{2}}^{\mathsf T}\,,
\end{equation}
while $ -\mathbf F $  acts on the wheel with opposite sense.
The control input $u$ is used for balancing and steering the unicycle.

The equations of motion of the unicycle are derived in  Appendix~\ref{app:unicycle_model_deriv} in a similar way as for the rolling wheel;
the differences between the two models are highlighted below.
The rolling constraints of the wheel still hold, yielding two kinematic constraints \eqref{eq:kin_constr} and one geometric constraint \eqref{eq:geom_constr_eq_z}.
Thus, the unicycle is a nonholonomic mechanical system with ${n_{\rm g}=3}$ geometric constraints 
and ${n_{\rm k}=2}$ kinematic constraints. 
Accordingly, ${n_q=9-n_{\rm g}=6}$ generalized coordinates 
have to be chosen to describe the system unambiguously; let these be
\begin{equation}
	\begin{split}
		\big(
		x_{\rm G}
		,\ y_{\rm G} 
		,\ \vartheta 
		,\ \psi 
		,\ \varphi
		,\ r 
		\big)%
		\,.
	\end{split}
\end{equation}
Moreover, ${n_\sigma = n_q - n_{\rm k} = 4}$ pseudovelocities have to be chosen; let these be the components of the angular velocity  $\bm\omega$ of the wheel (see \eqref{eq:omega} in Appendix~\ref{app:unicycle_model_deriv}) and the axle directional velocity component of the point mass:
\begin{align}\label{eq:pseudoveloc_defs_unicycle}
	\def\arraystretch{1.0}
	\setlength{\arraycolsep}{0pt}
	\begin{array}{rlrlrl}
		\SIGMA_1(t)        &:= \dot\vartheta
		,\quad 
		&\SIGMA_2(t)       &:= \dot\psi\sin\vartheta + \dot\varphi
		,\quad 
		\\
		\SIGMA_3(t)        &:= \dot\psi\cos\vartheta
		,\quad 
		&\SIGMAR(t)        &:=\dot r - \dot\vartheta R. 
	\end{array}
\end{align}

The Appellian approach yields the equations of motion
\begingroup
\allowdisplaybreaks
\begin{align}
	\label{eq:eqmot_unicycle}
	&\begin{cases}
		\begin{aligned}
			\dot \omega_{1} &= \frac{1}{5 m R^2 + 4 m_{0} r^{2}}\big(
			- 4 \omega_{1}^{2}  m_{0} R  r 
			-\omega_{3}^{2}( m R^2 
			+ 4 m_{0} r^{2} )\tan{\vartheta\,} 
			\\&\qquad
			- 8  \omega_{1} \sigma m_{0} r
			+ \omega_{2} \omega_{3} (6 m R^2  
			+ 4 m_{0} R  r  \tan{\vartheta\,} )
			\\&\qquad
			- 4 m_{0} g r \cos{\vartheta\,} 
			+ 4 m g R \sin{\vartheta\,} 
			+ 4 R u
			\big), 
			\\
			\dot \omega_{2} &= \frac{2 }{3 m R^2 + 2 m_{0} R^2  + 12 m_{0} r^{2}}\big(
			- 2 \omega_{1} \omega_{2}  m_{0} R  r 
			\\&\qquad
			+ \omega_{1} \omega_{3}  (m_{0} R^2 - m R^2 - 4 m_{0} r^{2})  
			+ 2 \omega_{3} \sigma  m_{0} R  
			\big), 
			\\
			\dot \omega_{3} &= \frac{1}{3 m R^2 + 2 m_{0} R^2  + 12 m_{0} r^{2}}\big(
			- \omega_{1} \omega_{2} ( 6 m R^2 + 4 m_{0} R^2 ) 
			\\&\qquad
			+ \omega_{1} \omega_{3} \big((3 m R^2 + 2 m_{0} R^2  + 12 m_{0} r^{2}) \tan{\vartheta\,}
			\big) 
			\\&\qquad
			- 24 \omega_{3} \sigma  m_{0} r 
			- 20 m_{0} R  r
			\big), 
			\\
			\dot \vartheta &= \omega_{1}, 
			\\
			\dot \sigma &%
			= 
			\omega_{1}^{2}  r 
			+ \omega_{3}^{2} r 
			- \omega_{2} \omega_{3} R 
			- g \sin{\vartheta\,} 
			+ \frac{1}{m_{0}}u ,
			\\
			\dot r &%
			= \sigma + \omega_{1} R, 
		\end{aligned}
	\end{cases}\nonumber
	\\
	&\begin{cases}
		\begin{aligned}
			\dot \psi =&\, \SIGMA_{3} \frac{1}{\cos{\vartheta}}
			\,,\\[1ex]
			\dot \varphi =&\, \SIGMA_{2} -  \SIGMA_{3} \tan{\vartheta} \, 
			\,,\\
			\dot x_{\rm G} =&\, \SIGMA_{1} R \sin{\psi} \cos{\vartheta}  + \SIGMA_{2} R \cos{\psi}
			\,,\\
			\dot y_{\rm G} =&\, - \SIGMA_{1} R  \cos{\psi} \cos{\vartheta}  + \SIGMA_{2} R  \sin{\psi} \,,
		\end{aligned}
	\end{cases}
\end{align}
\endgroup
which is a ${n = 9 - n_{\rm g} -n_{\rm k}/2 = 5}$\,DoF nonholonomic mechanical system. 
The equations  of motion \eqref{eq:eqmot_unicycle} of the unicycle can be divided into essential dynamics (first six equations) and hidden dynamics (remaining four equations) \cite{Routh_1884}.
Observe that the hidden dynamics of the unicycle \eqref{eq:eqmot_unicycle} are identical to that of the rolling wheel \eqref{eq:eq_mot_disc}. 
Moreover, considering ${m_0 = 0}$ and ${u = 0}$ the essential dynamics in \eqref{eq:eqmot_unicycle} simplify to that in \eqref{eq:eq_mot_disc}.
The dynamical model \eqref{eq:eqmot_unicycle} of the unicycle is formulated as a control affine system
${\dot{\mathbf x} = f(\mathbf x) + g(\mathbf x)u}$
with the state vector 
\begin{equation} \label{eq:unicycle_state_vector}
	\mathbf x
	=
	\left[\begin{array}{cccccc;{2pt/2pt}cccc}
		\SIGMA_{1}
		&\SIGMA_{2}
		&\SIGMA_{3}
		&\vartheta
		&\SIGMAR
		&r
		&\psi
		&\varphi
		&x_{\rm G}
		&y_{\rm G}
	\end{array}\right]^{\mathsf T}\!,
\end{equation}
and control input $u$; so this model is ready for control design.

\subsection{Steady states} 
\label{subsec:unicycle_steady_states}

Considering ${u \equiv 0}$ (i.e., the point mass freely moves along the axle), the unicycle model \eqref{eq:eqmot_unicycle} possesses the steady state motion with essential dynamics 
\begin{align}
	\label{eq:essential_equilibrium_consts_unicycle}
	\def\arraystretch{1.5}
	\setlength{\arraycolsep}{0pt}
	\begin{array}{rlrlrl}
		\SIGMA_1(t)       &\,\equiv   \SIGMA_{1*},\quad 
		&\SIGMA_2(t)       &\,\equiv   \SIGMA_{2*},\quad
		&\SIGMA_3(t)       &\,\equiv   \SIGMA_{3*},
		\\
		\vartheta(t)       &\,\equiv  \vartheta_{*},
		&\SIGMAR(t)        &\,\equiv  \SIGMARS,     
		&r(t)              &\,\equiv  r_{*}. 
	\end{array}
\end{align}
Substituting this into the first six equations of~\eqref{eq:eqmot_unicycle} yields 
\begin{align}
	\label{eq:steadystaterelation_pseudoveloc_unicycle}
	\begin{split}
		&\SIGMA_{2*} \SIGMA_{3*} \big(6 m R^2 + 4 m_{0} R  r_{*} \tan{\vartheta_{*} }\big) 		
		- 4 m_{0} g r_{*} \cos{\vartheta_{*} } 
		\\&\quad
		- \SIGMA_{3*}^{2} \big(m R^2 + 4 m_{0} r_{*}^{2} \big) \tan{\vartheta_{*} }
		+ 4 m g R \sin{\vartheta_{*}} = 0\,,
		\\[1ex]
		&
		\SIGMA_{3*}^{2} r_{*} - \SIGMA_{2*} \SIGMA_{3*} R  - g \sin{\vartheta_{*} } = 0\,,
	\end{split}
\end{align}
while the hidden motion of the unicycle is the same as the hidden motion of the rolling wheel described in~\eqref{eq:const_gen_velocs_disc} and~\eqref{eq:steadystate_hiddendyn_disc}.

Using \eqref{eq:pseudoveloc_defs_unicycle}, the relations \eqref{eq:steadystaterelation_pseudoveloc_unicycle} can be reformulated using the generalized velocities
($\dot \psi_{*}$, $\dot \vartheta_{*}$, $\dot \varphi_{*}$,  $\dot r_{*}$)%
:
\begin{align}
	\label{eq:steadystaterelation_genveloc_unicycle}
	\begin{split}
		& \dot\psi_{*}^{2} \big(4 m_{0} R  r_{*} \sin^{2}{\vartheta_{*} } + (5 m R^2 - 4 m_{0} r_{*}^{2}) \sin{\vartheta_{*} } \cos{\vartheta_{*} }\big) 
		\\&\quad 
		+ \dot\psi_{*} \dot\varphi_{*} \big(6 m R^2 \cos{\vartheta_{*} } + 4 m_{0} R  r_{*} \sin{\vartheta_{*} }\big) 
		\\&\quad 
		- 4 m_{0} g r_{*} \cos{\vartheta_{*} } + m g R \sin{\vartheta_{*} } = 0\,,
		\\[1ex]
		& \dot\psi_{*}^{2} \big(r_{*} \cos^{2}{\vartheta_{*} } - R \sin{\vartheta_{*} } \cos{\vartheta_{*} } \big) 
		\\&\quad 	
		- \dot\psi_{*} \dot\varphi_{*} R  \cos{\vartheta_{*} }  
		- g \sin{\vartheta_{*} } = 0\,,
	\end{split}
\end{align}
Then one may express the pitch rate $\dot{\varphi}_{*}$ and point mass position $r_{*}$ as a function of the tilt angle $\vartheta_{*}$ and the yaw rate $\dot{\psi}_{*}$:
\begin{align}
	\label{eq:steadystatesurfaces_unicycle}
	\begin{split}
		\dot\varphi_{*} &= \frac{\tan{\vartheta_{*} }}{ \dot\psi_{*} R (6 \dot\psi_{*}^{2} m R  \cos^{3}{\vartheta_{*}} - 4 m_0 g) }
		\big(
		- 5 \dot\psi_{*}^{4} m  R^{2} \cos^{4}{\vartheta_{*} } 
		\\&\qquad         
		+ 4 \dot\psi_{*}^{2} g R  ( m_{0} \cos{\vartheta_{*} }  	
		- m  \cos^{3}{\vartheta_{*} }  
		)
		+4 m_{0}  g^{2} 
		\big) 
		\,,
		\\[1ex]
		r_{*} &= \frac{mR (\dot\psi_{*}^{2} R  \cos{\vartheta_{*}} + 2 g )\sin{\vartheta_{*}} \cos{\vartheta_{*}} }
		{6 \dot\psi_{*}^{2} m R  \cos^{3}{\vartheta_{*}} - 4 m_0 g }\,,
	\end{split}
\end{align}
if 
\begin{align}
	\label{eq:steadystates_exceptions_unicycle}
	\begin{split}
		\dot{\psi}_{*} &\neq 0\,,
		\quad
		\dot{\psi}_{*} \neq \pm\sqrt{\dfrac{2 m_0 g}{3 m R  \cos^{3}{\vartheta_{*}}}},
	\end{split}
\end{align}
holds. 
For the non-generic cases, when~\eqref{eq:steadystates_exceptions_unicycle} does not hold, the steady state motions 
are explained below.

When the yaw rate is zero ${ \dot{\psi}_{*} = 0}$, straight rolling occurs and~\eqref{eq:steadystaterelation_genveloc_unicycle} simplify to  
\begin{align}
	\label{eq:straight_rolling_unicycle}
	\begin{split}
		&{4 m R \sin{\vartheta_{*} }} - {4 m_0 r_{*} \cos{\vartheta_{*} }} = 0\,,
		\quad 
		g\sin\vartheta_{*} = 0\,.
	\end{split}
\end{align}
These are only satisfied for zero tilt angle ${\vartheta_{*}=0}$ and centered point mass ${r_{*}=0}$, independent of the pitch rate $\dot{\varphi}_{*}$. 

In case of ${\dot{\psi}_{*} = \pm\sqrt{{2 m_0 g}/({3 m R  \cos^{3}{\vartheta_{*}}})}}$, \eqref{eq:steadystaterelation_genveloc_unicycle} still results in zero tilt angle ${\vartheta_{*}=0}$ and 
\begin{align}
	\label{eq:nontilted_turning_unicycle}
	\begin{split}
		\dot{\psi}_{*1,2} &= \pm \sqrt{\frac{{2} {m_0} {g} }{3 {m} {R} }}\,,
		\quad
		\dot{\varphi}_{*1,2} = \pm  \sqrt{\dfrac{{2} {m_0} {g} }{3 m R^{3}}} r_{*}\,,
	\end{split}
\end{align}            
which describe a turning-rolling steady state motion with a non-tilted wheel. 
This is a new behavior of the unicycle which does not exist for rolling wheel, as the latter must be tilted in order to have a turning-rolling steady state motion, cf.~\eqref{eq:steadystaterelation_genveloc_disc}.
Note that, this steady state can only occur for the specific yaw rates ${\dot{\psi}_{*1,2}}$ depending only on physical parameters, while the corresponding pitch rates ${\dot{\varphi}_{*1,2}}$ also depend only on the point mass position $r_{*}$.
The non-tilted turning steady state becomes a spinning steady state (${\dot{\varphi}_{*1,2}}=0$) if the point mass is at the center ${r_{*}=0}$.

The other spinning steady state can be obtained by substituting zero pitch rate ${\dot{\varphi}_{*}=0}$ into  \eqref{eq:steadystaterelation_genveloc_unicycle}, which yields
\begin{align}
	\label{eq:spinning_steadystate_unicycle}
	\begin{split}
		&\dot\psi_{*}^{2} \big(
		(5 m R^{2} -  4 m_0 r_{*}^{2} )\sin{\vartheta_{*} } \cos{\vartheta_{*} } 
		+ {4 m_0 R  r_{*} \sin^{2}{\vartheta_{*} }}  
		\big)  
		\\&\quad
		+ {4 m g R \sin{\vartheta_{*} }}  
		-  {4 m_0 g  r_{*} \cos{\vartheta_{*} }} 
		= 0\,,
		\\[1ex]
		& \dot\psi_{*}^{2} \big(  r_{*} \cos^{2}{\vartheta_{*} } - R \sin{\vartheta_{*} } \cos{\vartheta_{*} } \big) - g \sin{\vartheta_{*} } = 0\,.
	\end{split}
\end{align}
Similar to the spinning wheel, such steady state motion exists for zero tilt angle (${\vartheta_{*}=0}$) when the point mass is at the center ${r_{*}=0}$. 
However, in case of the unicycle, 
spinning type steady state motion also exists with non-zero tilt angle  (${\vartheta_{*}\neq 0}$) when
\begin{align}
	\label{eq:tilted_spinning_steadystate_unicycle}
	\dot{\psi}_{*} &= \pm\sqrt{
		\frac{
			4 m_0 g   r_{*} \cos{\vartheta_{*} } - 4 m g R \sin{\vartheta_{*} } 
		}{
			(5 m R^{2}  - 4 m_0 r_{*}^{2}) \sin{\vartheta_{*} } \cos{\vartheta_{*} } + 4 m_0 R r_{*} \sin^{2}{\vartheta_{*} }
		}
	}\,,
	\nonumber\\[1ex]
	r_{*} &= \frac{R  \tan{\vartheta_{*} }}{2 m_0} \Big(m \cos^{2}{\vartheta_{*} } + m_0 
	\\&\qquad\qquad\quad\nonumber
	+ \sqrt{m^{2} \cos^{4}{\vartheta_{*} } + 3 m m_0 \cos^{2}{\vartheta_{*} } + m_0^{2}}\Big)\,.
\end{align}
We refer to this as tilted spinning,
which is also a new behavior of the unicycle compared to the rolling wheel.
According to~\eqref{eq:tilted_spinning_steadystate_unicycle}, the point mass is always above the wheel center since ${{\rm sgn}\,r_{*} = {\rm sgn}\,\vartheta_{*}}$ or ${r_{*}\,\vartheta_{*} > 0}$; cf.~Figure~\ref{fig:unicycle_mech_model}.
Investigating special cases like ${\vartheta_{*}=0}$ or ${r_{*}=0}$ leads to steady states already discussed above.

One limiting factor of our unicycle model is that the point mass may slide below the ground which is physically not feasible.
To exclude these cases, one can calculate the $z$ coordinate of the position vector $\mathbf r_{\rm A} = \mathbf r_{\rm G} + \mathbf r_{\rm GA}$ in frame ${\rm F_0}$  and require it to be positive.
This yields
${
	R \cos{\vartheta_{*}} + r_{*} \sin{\vartheta_{*}} > 0\,.
}$

\begin{figure}[t!]
	\centering
	\includegraphics[width=1\columnwidth]{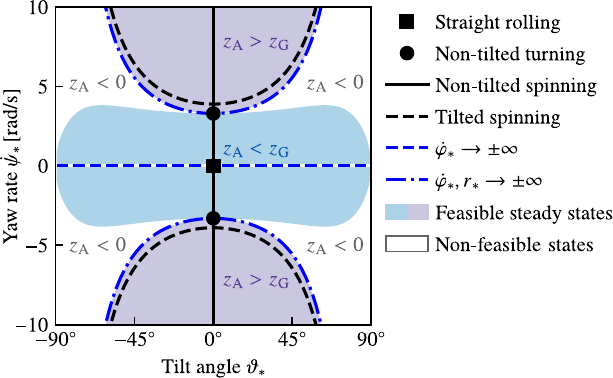}
	\caption{Summary of the steady state motions of the unicycle}
	\label{fig:steadystate_summary_unicycle}
	\vspace{-4mm}
\end{figure}

All steady state motions of the unicycle are summarized in Figure~\ref{fig:steadystate_summary_unicycle} in the plane of the steady state tilt angle $\vartheta_{*}$ and the yaw rate ${\dot{\psi}_{*}}$.  
The straight rolling, i.e.,~${\vartheta_{*}=0}$ and ${\dot{\psi}_{*}=0}$, is marked by a black square~$\blacksquare$; 
in this case the steady state pitch rate $\dot{\varphi}_{*}$ may be arbitrary and the point mass is at wheel center ${r_{*}=0}$. 
The non-tilted turning, i.e.,~${\vartheta_{*}=0}$ and ${\dot{\psi}_{*}=\dot{\psi}_{*1,2}}$, 
is marked by black dots~{\small $\CIRCLE$};
in this case the pitch rate $\dot{\varphi}_{*}$ and the point mass position $r_{*}$  linked as in~\eqref{eq:nontilted_turning_unicycle}.
The regular (non-tilted) spinning steady state motions, i.e.,~${\vartheta_{*}=0}$ and ${\dot{\varphi}_{*}=0}$, are shown by the solid black line (\textcolor[RGB]{0,0,0}{\bfseries ---}); 
in this case the yaw rate $\dot{\psi}_{*}$ may be arbitrary and the point mass is at the wheel center ${r_{*}=0}$.
The tilted spinning steady states, 
i.e.,~${\dot{\varphi}_{*}=0}$ and ${\vartheta_{*}\neq 0}$, 
are shown by dashed black lines (\textcolor[RGB]{0,0,0}{\bfseries -- --}); 
the corresponding yaw rate $\dot{\psi}_{*}$ and point mass position $r_{*}$ are expressed by~\eqref{eq:tilted_spinning_steadystate_unicycle}.
The general turning-rolling steady states are divided into separate areas.
The light blue \textcolor[RGB]{172,210,231}{$\blacksquare\!\blacksquare$} and light purple \textcolor[RGB]{201,199,223}{$\blacksquare\!\blacksquare$} areas show the physically feasible steady states when the point mass is placed below ${0<z_{\rm A}<z_{\rm G}}$ or above ${z_{\rm A}>z_{\rm G}}$ the wheel center point~$\rm G$, respectively.
The white area {$\bm\sqsubset \! \bm\sqsupset$} represents the physically unfeasible steady states when the point mass is at or below the ground level, that is,  ${z_{\rm A}\leq 0}$.

Note that Figure~\ref{fig:steadystate_summary_unicycle} was constructed using the numerical parameters  in Table~\ref{tab:physical_consts_disc}. 
Using different parameters, like different mass ratios ${m_0/m}$, the main structure of the steady states remain qualitatively similar, but the shape and size of the parameter regions related to the non-feasible steady states may be  different.

The steady states have been categorized with considering zero input force (${u \equiv 0}$). 
Assuming a nonzero constant input (${u\equiv u_*}$) may result in further steady states.
For example, a straight rolling case may occur with tilted wheel when the mass is held `above' the wheel center. 
Studying such steady states are out of the scope of the present study.

\subsection{Stability of steady states}

\begin{figure*}[p]
	\centering
	\includegraphics[width=0.95\columnwidth]{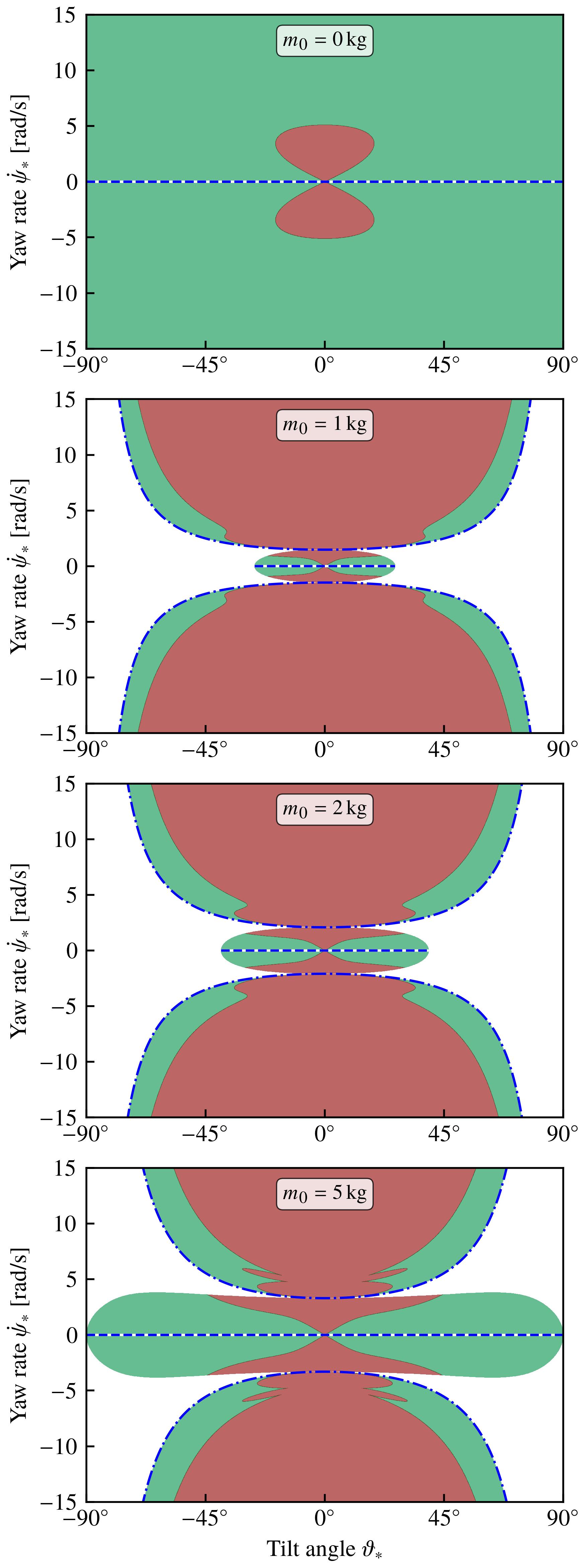}
	\includegraphics[width=0.95\columnwidth]{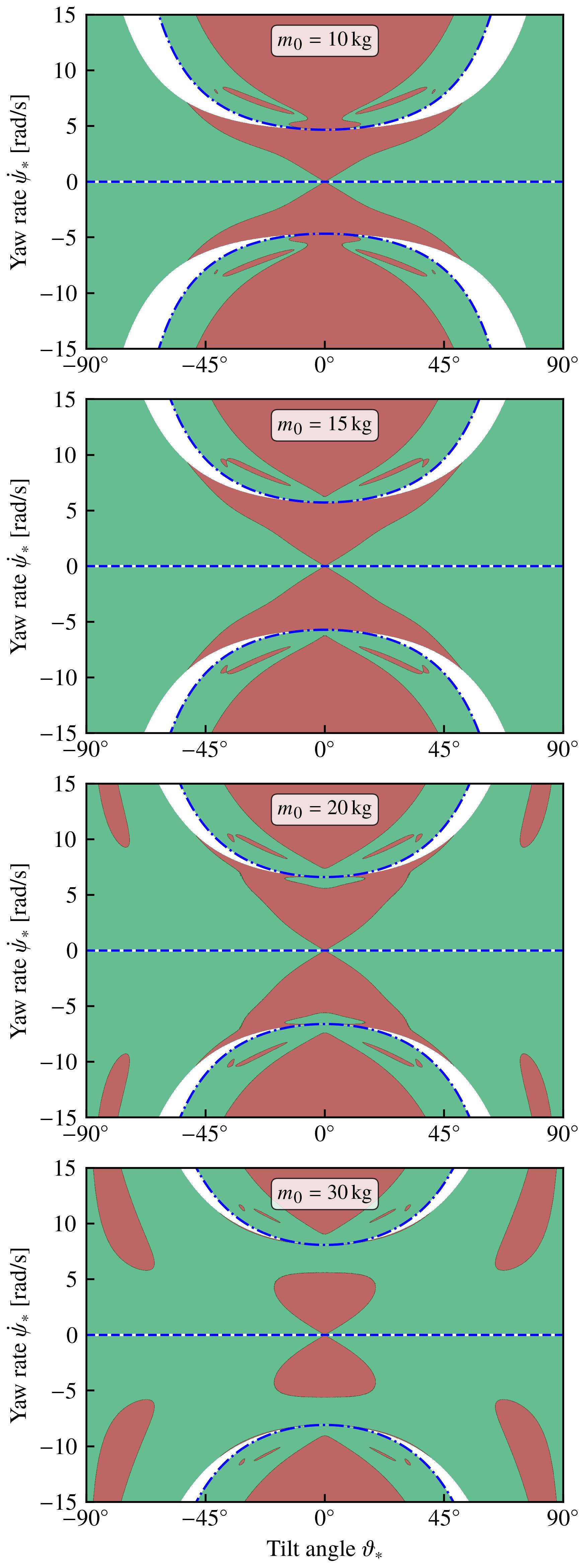}
	\caption{
		Stability of steady state motions of the unicycle for various $m_{0}$ values: stable and unstable steady states are denoted by green and red, respectively, while white denotes the unfeasible steady states when the point mass hits the ground.
	}
	\label{fig:stability_unicycle}
\end{figure*}

Linear stability analysis is performed to determine the stability of the previously discussed steady state motions of the unicycle.
Linearizing \eqref{eq:eqmot_unicycle} around a steady state $\mathbf x_{*}$ leads to the form
${\dot{\tilde{\mathbf x}} =   \mathbf A \tilde{\mathbf x}}$
where ${\tilde{\mathbf x} = \mathbf x - \mathbf x_{*}}$ denotes the perturbed states and state matrix ${\mathbf{A} = f'(\mathbf{x}_{*})}$ can be written as
\begin{equation}
	\label{eq:Amat_rollturn_unicycle}
	\arraycolsep=3.2pt
	\mathbf A =     \left[\begin{array}{cccccc;{2pt/2pt}cccc}%
		0       & A_{12}  & A_{13} & A_{14} & 0      & A_{16} & 0       & 0 & 0 & 0 \\
		A_{21}  & 0       & 0      & 0      & A_{25} & 0      & 0       & 0 & 0 & 0 \\
		A_{31}  & 0       & 0      & 0      & A_{35} & 0      & 0       & 0 & 0 & 0 \\
		1       & 0       & 0      & 0      & 0      & 0      & 0       & 0 & 0 & 0 \\
		0       & A_{52}  & A_{53} & A_{54} & 0      & A_{56} & 0       & 0 & 0 & 0 \\
		A_{61}  & 0       & 0      & 0      & 1      & 0      & 0       & 0 & 0 & 0 \\\hdashline[2pt/2pt]
		0       & 0       & A_{73} & A_{74} & 0      & 0      & 0       & 0 & 0 & 0 \\
		0       & 1       & A_{83} & A_{84} & 0      & 0      & 0       & 0 & 0 & 0 \\
		A_{91}  & A_{92}  & 0      & 0      & 0      & 0      & A_{97}  & 0 & 0 & 0 \\
		A_{101} & A_{102} & 0      & 0      & 0      & 0      & A_{107} & 0 & 0 & 0
	\end{array}\right],
\end{equation}
whose elements are given in Appendix~\ref{app:Amatgeneral_elements_unicycle}. 
Note that the steady state tilt angle $\vartheta_{*}$, yaw rate $\dot{\psi}_{*}$, pitch rate $\dot{\varphi}_{*}$ and point mass position $r_{*}$ in the elements of $\mathbf A$ are not independent of each other, they must satisfy \eqref{eq:steadystaterelation_genveloc_unicycle}.
Therefore, each of the different steady state motions (turning-rolling, straight rolling, non-tilted turning) must be analyzed separately.
In this paper we focus on the straight rolling and turning-rolling motions while the non-tilted turning is left for future research.
Also note that, the spinning steady state is a special case of the more general turning-rolling steady state; since it has minor practical relevance its further analysis is omitted.

	The characteristic equation  ${\det(\lambda \mathbf I - \mathbf{\hat{A}}) = 0}$, where $\mathbf{\hat{A}}$ is the top left 6 by 6 part of $\mathbf A$ in \eqref{eq:Amat_rollturn_unicycle}, takes the form 
\begin{equation}
	\begin{split}
		\label{eq:chp_general_unicycle}
		& \big(\lambda^{4}
		- \lambda^{2} (A_{12} A_{21} + A_{13} A_{31} + A_{14} + A_{16} A_{61} + A_{25} A_{52}
		\\&
		+ A_{35} A_{53} + A_{56} )
		+ A_{12} A_{21} A_{35} A_{53} 
		- A_{12} A_{25} A_{31} A_{53} 
		\\&
		- A_{12} A_{25} A_{56} A_{61} 
		- A_{13} A_{21} A_{35} A_{52} 
		+ A_{13} A_{25} A_{31} A_{52} 
		\\&
		- A_{13} A_{35} A_{56} A_{61} 
		+ A_{16} A_{25} A_{52} A_{61} 
		+ A_{16} A_{35} A_{53} A_{61} 
		\\&
		+ A_{12} A_{21} A_{56} 
		- A_{12} A_{25} A_{54} 
		+ A_{13} A_{31} A_{56} 
		- A_{13} A_{35} A_{54} 
		\\&
		+ A_{14} A_{25} A_{52} 
		+ A_{14} A_{35} A_{53} 
		- A_{16} A_{21} A_{52} 
		- A_{16} A_{31} A_{53} 
		\\&
		+ A_{14} A_{56} 
		- A_{16} A_{54}
		\big)
		{\lambda^{2} 		}
		=0\,.
	\end{split}
\end{equation}

\subsubsection{Turning-rolling steady state}

When calculating the characteristic roots of~\eqref{eq:chp_general_unicycle}  the steady state pitch rate $\dot{\varphi}_{*}$ and point mass position $r_{*}$ must be substituted according to~\eqref{eq:steadystatesurfaces_unicycle}.
Due to the highly complicated expressions, numerical evaluation is used to obtain the characteristic roots and determine the stability of the turning-rolling steady states.
The results are shown in Figure~\ref{fig:stability_unicycle} for several mass ratios with ${m_0 = 0, 1, 2, 5,10,15,20,30}$\,kg, while the other physical parameters are presented in Table~\ref{tab:physical_consts_disc}.
The same coloring scheme is used as in Fig.~\ref{fig:levels_and_stability_disc}, namely, green (\textcolor[RGB]{102,191,146}{$\blacksquare\!\blacksquare$}) and red (\textcolor[RGB]{191,102,102}{$\blacksquare\!\blacksquare$}) areas represent stable and unstable steady state solutions, respectively, 
while the white area ({$\bm\sqsubset \!\!\! \bm\sqsupset$}) represents the physically unfeasible steady states with point mass at or below the ground level, cf.~Fig.~\ref{fig:steadystate_summary_unicycle}.
The case with \({m_0=0}\) is identical to the rolling wheel in Figure~\ref{fig:levels_and_stability_disc}(c). 
By the addition the point mass \({m_0>0}\), the stability properties change drastically since the number of degrees of freedom is increased, additional types of equilibria are born (the tilted spinning and the vertical turning), the point mass may be above or below the wheel, and it even may hit the ground, as was discussed in Section~\ref{subsec:unicycle_steady_states} and Figure~\ref{fig:steadystate_summary_unicycle}.

With small point masses (such as \({m_0=1, 2}\)\,kg) the vertical spinning (\({\vartheta_{*}=0}\)) becomes unstable as the point mass is pushed away from the spinning wheel by the centrifugal force. 
With large point masses (such as \({m_0=15, 20, 30}\)\,kg) a stable yaw rate \(\dot \psi_{*} \) band appears, but the spinning steady states remains unstable for large enough yaw rates.

\subsubsection{Straight rolling}

The steady state is given as
\eqref{eq:straight_rolling} and all the other states become zero,
considering that the initial values $x_{\rm 0}$, $y_{\rm 0}$ and $\dot{\psi}_0$ are zeros. 
The state matrix \eqref{eq:Amat_rollturn_unicycle} simplifies to
\begin{equation}
	\label{eq:Amat_straight_unicycle}
	\arraycolsep=3.2pt
	\mathbf A =
	\left[\begin{array}{cccccc;{2pt/2pt}cccc}%
		0 & 0 & \frac{6 }{5}\dot\varphi_{*} & \frac{4 g}{5 R} & 0 & - \frac{4 m_{0} g}{5 m R^2} & 0 & 0 & 0 & 0
		\\
		0 & 0 & 0 & 0 & 0 & 0 & 0 & 0 & 0 & 0
		\\
		- 2 \dot\varphi_{*} & 0 & 0 & 0 & 0 & 0 & 0 & 0 & 0 & 0
		\\
		1 & 0 & 0 & 0 & 0 & 0 & 0 & 0 & 0 & 0
		\\
		0 & 0 & - R \dot\varphi_{*} & - g & 0 & 0 & 0 & 0 & 0 & 0
		\\
		R & 0 & 0 & 0 & 1 & 0 & 0 & 0 & 0 & 0
		\\\hdashline[2pt/2pt]
		0 & 0 & 1 & 0 & 0 & 0 & 0 & 0 & 0 & 0
		\\
		0 & 1 & 0 & 0 & 0 & 0 & 0 & 0 & 0 & 0
		\\
		0 & R & 0 & 0 & 0 & 0 & 0 & 0 & 0 & 0
		\\
		- R & 0 & 0 & 0 & 0 & 0 & R \dot\varphi_{*} & 0 & 0 & 0
	\end{array}\right].
\end{equation}
When comparing this with \eqref{eq:Amat_straight_disc} obtained for the rolling wheel, there are two extra rows and columns, 5 and 6, related to the new states $r$ and $\SIGMAR$.

\begin{figure}[b!]
	\centering
	\includegraphics[width=0.95\columnwidth]{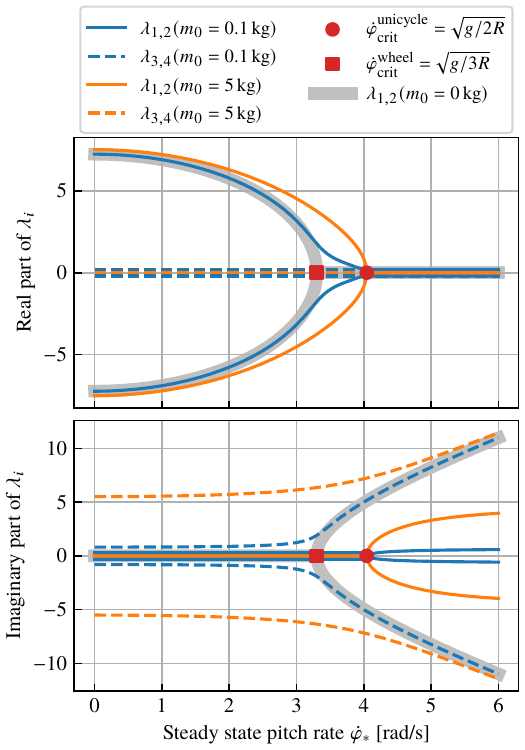}
	\caption{Characteristic roots as a funtion the steady state pitch rate $\dot\varphi_{*}$ for straight rolling comparing the cases of rolling wheel and unicycle}
	\label{fig:charroots_unic}
\end{figure}

The characteristic polynomial \eqref{eq:chp_general_unicycle} reduces to
\begin{equation}
	\begin{split}
		\label{eq:chp_straight_unicycle}
		\big( a \lambda^{4} + b \lambda^{2} + c \big)
		{\lambda^{2}}
		=0\,,
	\end{split}
\end{equation}
where
\begin{equation}
	\begin{split}
		a &= {5 m R^{2} },\,
		\\
		b &= {12 \dot\varphi_{*}^{2} m R^{2}  + 4 ( m_0  - m) g R  },\,
		\\
		c &= 4 m_0 g ( {2 \dot\varphi_{*}^{2} R  -  g  } ).
	\end{split}
\end{equation}
Applying the same arguments as in Appendix~\ref{app:disc_stability}, 
the straight rolling is said to be stable if the coefficients $a,b$ and $c$ have the same signs. 
\color{black}
While $a>0$ always holds, $b>0$ and  $c>0$ lead to
\begin{equation}
	|\dot\varphi_{*}| > \sqrt{\frac{ ( m  - m_{0} )  g}{3 m R}} \quad \text{ and } \quad |\dot\varphi_{*}| > \sqrt{\frac{g}{2 R}}\,,
\end{equation}
respectively. 
The second condition is always stronger, so the critical pitch rate of the unicycle is 
\begin{equation}\label{eq:critpitchrate_unicycle}
	\dot\varphi_{\rm crit} = \sqrt{\frac{g}{2 R}}\,.
\end{equation}
That is, the straight rolling of the uncontrolled unicycle is stable if ${ |\dot\varphi_{*}| > \dot{\varphi}_{\rm crit} }$; it is unstable if ${ |\dot\varphi_{*}| < \dot{\varphi}_{\rm crit} }$.

Note that, the critical pitch rate~\eqref{eq:critpitchrate_unicycle} is independent of the masses $m$ and $m_0$. 
Also, it is larger than the critical pitch rate of the rolling disc \eqref{eq:critpitchrate_disc}.
The dependency of the characteristic roots on the steady state pitch rate $ \dot\varphi_{*} $ are summarized in Figure~\ref{fig:charroots_unic} for both the rolling wheel (${m_0 = 0}$) and the unicycle (${m_0 > 0}$).
Observe that in case of the rolling wheel the characteristic roots $\lambda_{1,2}$ (thick gray curves) determine the criticality. 
Contrarily, in case of the unicycle (even for small values of $m_0$!), the criticality is determined by the characteristic roots $\lambda_{3,4}$ (orange curves) rather than by the characteristic roots $\lambda_{1,2}$ (blue curves).
This resolves the seemingly contradictory issue that the critical pitch rate `jumps' from \eqref{eq:critpitchrate_disc} to \eqref{eq:critpitchrate_unicycle} when $m_0$ becomes positive.
On the other hand, this is not surprising in view of the different degrees of freedom in the two systems.%

\subsubsection{Spinning on the spot}

The stability analysis of this steady state is covered by the analysis of the turning-rolling motion.
Since the motion has minor practical relevance for unicycles further discussions are omitted here.

\section{Steering control of the unicycle}\label{sec:unic_control}

The straight rolling steady state~\eqref{eq:straight_rolling} is used as the basis for designing the steering controller of the unicycle. 
For simplicity, it is assumed that the unicycle has an initial pitch rate ${ \dot{\varphi}_0 = \dot{\varphi}_{*} }$ and initially it travels along the $x_0$ axis.
Above we showed that the straight rolling steady state might be unstable or stable without control.
Here we demonstrate that this motion can be made asymptotically stable with appropriate feedback control.

The linear state space model of the unicycle assumes the form 
${ \dot{\tilde{\mathbf x}} = \mathbf A \tilde{\mathbf x} + \mathbf B  u }$
where ${\tilde{\mathbf x} = \mathbf x - \mathbf x_{*}}$.
The state matrix ${\mathbf{A} = f'(\mathbf{x}_{*})}$ is given in~\eqref{eq:Amat_straight_unicycle} and it depends on the steady state pitch rate $ \dot{\varphi}_{*} $. 
The input matrix ${\mathbf{B} = g(\mathbf{x}_{*})}$ is given by%
\begin{equation}\label{eq:Bmat_straight_unicycle}
	\begin{split}
		\mathbf B &= 
		\begin{bmatrix}
			\frac{4}{5 m R}
			\ \ 0
			\ \ 0
			\ \ 0
			\ \ \frac{1}{m_0}
			\ \ 0
			\ \ 0
			\ \ 0
			\ \ 0
			\ \ 0
		\end{bmatrix}^\mathsf T.
	\end{split}
\end{equation}

\subsection{Controllability}

The controllability matrix of the unicycle is obtained as
\begin{equation}
	\begin{split}
		\mathbf M_{\rm c} = \begin{bmatrix}
			\mathbf B 
			\ \ \mathbf A \mathbf B
			\ \ \mathbf A^{\!2} \mathbf B
			\ \ \dots
			\ \ \mathbf A^{\!9} \mathbf B
		\end{bmatrix}.
	\end{split}
\end{equation}
Substituting \eqref{eq:Amat_straight_unicycle} and \eqref{eq:Bmat_straight_unicycle} we obtain
\begin{equation}
	\label{eq:Mc_mat_struct}
	\setlength\arraycolsep{1.0pt}
	\mathbf M_{\rm c} = 
	\begin{bmatrix}
		M_{11} & 0       & M_{13} & 0       & M_{15} & 0       & M_{17} & 0       & M_{19} & 0        \\
		0      & 0       & 0      & 0       & 0      & 0       & 0      & 0       & 0      & 0        \\
		0      & cM_{42}  & 0      & cM_{44}  & 0      & cM_{46}  & 0      & cM_{48}  & 0      & cM_{410}  \\
		0      & M_{42}  & 0      & M_{44}  & 0      & M_{46}  & 0      & M_{48}  & 0      & M_{410}  \\
		M_{51} & 0       & M_{53} & 0       & M_{55} & 0       & M_{57} & 0       & M_{59} & 0        \\
		0      & M_{62}  & 0      & M_{64}  & 0      & M_{66}  & 0      & M_{68}  & 0      & M_{610}  \\
		0      & 0       & M_{73} & 0       & M_{75} & 0       & M_{77} & 0       & M_{79} & 0        \\
		0      & 0       & 0      & 0       & 0      & 0       & 0      & 0       & 0      & 0        \\
		0      & 0       & 0      & 0       & 0      & 0       & 0      & 0       & 0      & 0        \\
		0      & M_{102} & 0      & M_{104} & 0      & M_{106} & 0      & M_{108} & 0      & M_{1010}
	\end{bmatrix}\!,
\end{equation}
with ${c= -2\dot{\varphi}_{*}}$.
Notice that rows 2, 8 and 9 are full of zeros while rows 3 and 4 are linearly dependent. 
Thus, ${ \mathrm{rank}\, \mathbf{M}_{\rm c} = 6 }$ and the controllability matrix is not full row rank. 
That is,  at the linear level, the  unicycle is not fully controllable with the single input $u$.

When spelling out the 3\textsuperscript{rd} and 4\textsuperscript{th} equations in the linear system ${ \dot{\tilde{\mathbf x}} = \mathbf A \tilde{\mathbf x} + \mathbf B  u }$ one obtains
\begin{equation}
	\begin{split}
		\dot\SIGMA_{3} &= -2\dot{\varphi}_{*} \SIGMA_{1}\,,
		\qquad
		\dot\vartheta = \SIGMA_{1}\,
	\end{split}
\end{equation}
which lead to
\begin{equation}
	\label{eq:sigma3_theta_linear_relation_unicycle}
	\SIGMA_3 = -2\dot{\varphi}_{*}\vartheta\,\!.
\end{equation}
This means that the yaw rate $\dot\psi$ is linearly proportional to the tilt angle $\vartheta$, since $\SIGMA_3\approx \dot\psi$ for small tilt angles; cf.~\eqref{eq:pseudoveloc_defs_unicycle}. 
That is, to steer the unicycle it is necessary to tilt it accordingly. 

Note that one may steer the unicycle with a specific yaw rate $\dot{\psi}_{*}$ in~\eqref{eq:nontilted_turning_unicycle} even with zero tilt angle, but the model must be linearized around the non-tilted turning steady state of the unicycle. 
This is left for future research.

\subsection{Maneuvering}

In this study, two maneuvers are considered for the steering control of the unicycle: a lane change and a $ 90^\circ $ right turn, while assuming that the unicycle initially travels along the $x_0$ axis.
Considering the steady state velocity ${v_{\mathrm G*} = R \dot{\varphi}_{*} }$ of the center of gravity leads to the critical speed ${v_{\mathrm{G,crit}}=\sqrt{Rg/2}}$ 
for the critical pitch rate $\dot{\varphi}_{\mathrm{crit}}$ in \eqref{eq:critpitchrate_unicycle}.
For the parameters in Table~\ref{tab:physical_consts_disc}, we obtain
${v_{\mathrm{G,crit}}\approx 1.21}$\,m/s.
Thus, the lane change and right turn maneuvers are investigated for two speeds: ${v_{\mathrm G1}=1\,\mathrm{m/s}}$ which is below the critical speed, and ${v_{\mathrm G2}=5\,\mathrm{m/s}}$ which is above the critical speed.

When designing the controllers to execute these maneuvers, the non-reachable states $\omega_2$, $\omega_3$, $\varphi$, $x_{\rm G}$ are omitted;
cf.~\eqref{eq:unicycle_state_vector} and \eqref{eq:Mc_mat_struct}.
Then, according to the state matrix \eqref{eq:Amat_straight_unicycle} and input matrix \eqref{eq:Bmat_straight_unicycle} the reduced system becomes
\begin{equation}\label{eq:reduced_system}
	\begin{split}
		\dot\SIGMA_{1} &= \bigg( \frac{4g}{5R} - \frac{12}{5} \dot\varphi_*^2 \bigg)\vartheta - \frac{4 m_{0} g }{5 m R^2}r + \frac{4}{5 m R}u\,,
		\\
		\dot \vartheta &= \SIGMA_{1}\,,
		\\
		\DSIGMAR &=  \left(2  \dot\varphi_{*}^{2}  R- g\right) \vartheta  + \frac{1}{m_{0}}u\,,
		\\
		\dot r&= \SIGMAR +  R \omega_{1} \,,
		\\
		\dot \psi &= -2 \dot\varphi_* \vartheta\,\!,
		\\
		\dot y_{\rm G} &= - R \omega_1  + \dot\varphi_*  R \psi\,,
	\end{split}
\end{equation}
where the tildes above the states (representing perturbations) are omitted since the equilibrium values are zeros for all these states.
We formally define the states in this reduced system as outputs of the full linear system ${ \dot{\tilde{\mathbf x}} = \mathbf A \tilde{\mathbf x} + \mathbf B  u }$ when
designing the controllers for the lane change and turning maneuvers.

\subsubsection{Lane change}

For the lane change maneuver the state $y_{\rm G}$ can be utilized to plan the motion since the unicycle should move parallel to the $x_0$ axis at the beginning and at the end of the maneuver.
According to this, we consider the output
\begin{equation}\label{eq:output_lanechange}
	\mathbf y = \mathbf C \mathbf x := \begin{bmatrix}
		\SIGMA_{1} 
		&\vartheta 
		&\SIGMAR 
		&r 
		&\psi 
		&y_{\rm G}
	\end{bmatrix}^\mathsf T,
\end{equation}
(cf.~\eqref{eq:reduced_system}) with
\begin{equation}\label{eq:output_lanechange_matrix}
	\mathbf C = \begin{bmatrix}
		1 & 0 & 0 & 0 & 0 & 0 & 0 & 0 & 0 & 0\\
		0 & 0 & 0 & 1 & 0 & 0 & 0 & 0 & 0 & 0\\
		0 & 0 & 0 & 0 & 1 & 0 & 0 & 0 & 0 & 0\\
		0 & 0 & 0 & 0 & 0 & 1 & 0 & 0 & 0 & 0\\
		0 & 0 & 0 & 0 & 0 & 0 & 1 & 0 & 0 & 0\\
		0 & 0 & 0 & 0 & 0 & 0 & 0 & 0 & 0 & 1
	\end{bmatrix}.
\end{equation}
The output controllability matrix can be constructed as 
\begin{equation}
	\label{eq:contollability_matrix_output}
	\begin{split}
		\mathbf M_{\rm oc} = \begin{bmatrix}
			\mathbf C \mathbf B 
			\ \ \mathbf C \mathbf A \mathbf B
			\ \ \mathbf C \mathbf A^{\!2} \mathbf B
			\ \ \dots
			\ \ \mathbf C \mathbf A^{\!9} \mathbf B
		\end{bmatrix},
	\end{split}
\end{equation}
which has a maximal row rank, ${ \mathrm{rank}\,\mathbf M_{\rm oc} = 6 }$. 
That is, the unicycle is output controllable. 

Then, we apply the linear output feedback law 
\begin{equation}
	\label{eq:control_input_lanechange}
	u := - \mathbf K (\mathbf y - \mathbf y_{\rm des})\,,
\end{equation}
with control gains ${\mathbf K = [D_\vartheta\ P_\vartheta\ D_r\ P_{r}\ P_{\psi}\ P_{y}]}$ and 
\begin{equation}
	\label{eq:ref_traj_lanechange2}
	\arraycolsep=0pt
	\begin{array}{rcccccccccccl}
		\mathbf y_{\rm des} %
		=[
		&0
		&\hspace{1.0em}&0
		&\hspace{1.0em}&0
		&\hspace{1.0em}&0
		&\hspace{1.0em}&0
		&\hspace{1.0em}&y_{\rm des}(t)&]^\mathsf{T}\,,
	\end{array}  
\end{equation}
where $y_{\rm des}(t)$ encodes the desired trajectory in terms of $y_{\rm G}$.

\begin{figure*}[!t]
	\centering
	\includegraphics[width=0.8\columnwidth]{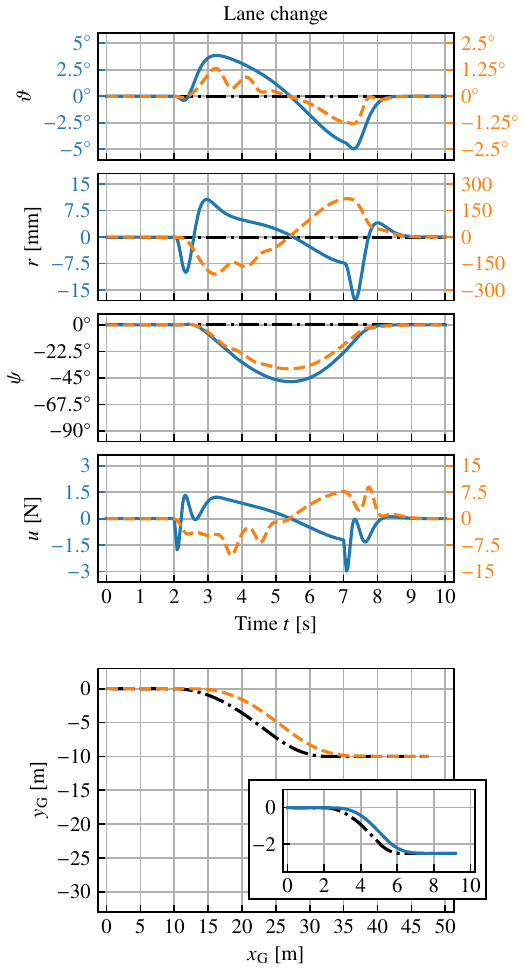}
	\hspace{2em}
	\includegraphics[width=0.8\columnwidth]{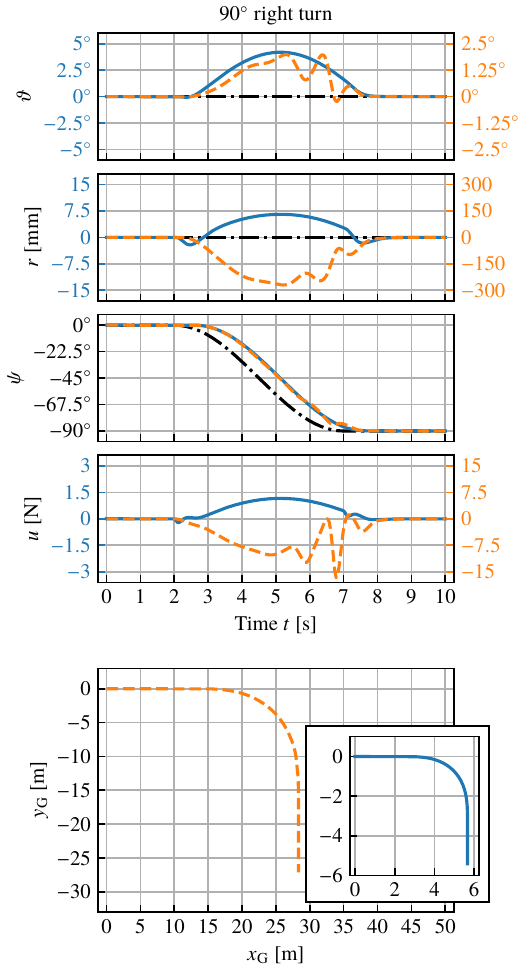}
	\\
	\includegraphics[width=1.5\columnwidth]{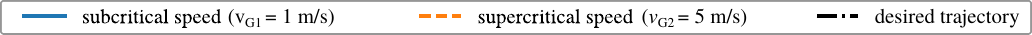}    
	\caption{Maneuvers: lane change (left column) and 90$^\circ$ right turn (right column) below the critical speed (solid blue) and above (dashed orange)}
	\label{fig:manuvers}
	\vspace{-5mm}
\end{figure*}

Here we consider 
\begin{equation}
	\label{eq:ref_traj_lanechange_y}
	\begin{split}
		y_{\rm des}(t) &= \begin{cases}
			\arraycolsep=0pt
			\begin{array}{lrl}
				0,               & 0\leq  &~t < 2,\\
				{\frac{\textcolor{black}{L}}{2}}\big(\cos\big(\tfrac{\pi}{5} (t-2)\big)-1\big),\hspace{1em} & 2\leq  &~t<7,\\
				-L,               & 7\leq &~t < 10,\\
			\end{array}
		\end{cases}
	\end{split}
\end{equation}
and we use ${L = 2.5\, {\rm m}}$ for speed ${v_{\mathrm G1}=1\,\mathrm{m/s}}$ and ${L = 10\, {\rm m}}$ for speed ${v_{\mathrm G2}=2\,\mathrm{m/s}}$.
The desired trajectories are depicted as a black dashed-dotted curves in the ${(x_0, y_0)}$-plane in the bottom left panel in Fig.~\ref{fig:manuvers}.
For simplicity, we consider the desired yaw angle ${{\psi_{\rm des}}(t)\equiv 0}$ in \eqref{eq:ref_traj_lanechange2}.
More sophisticated reference trajectories might be constructed utilizing \eqref{eq:steadystaterelation_genveloc_unicycle}.

Note that the proposed linear controller does not ensure the global stability of the nonlinear system. 
Nonlinear investigations are outside of the scope of the present study.
Also,  maneuvers with substantial non-zero $\dot\psi$ may make the unicycle deviate significantly from straight rolling steady state which may cause large deviation in the observed behavior.

The characteristic equation of the closed-loop system  is
\begin{equation}
	\det\big(\lambda \mathbf I - (\mathbf{A-BKC})\big)=0\,.
\end{equation}
By selecting the control gains in $\mathbf K$ in \eqref{eq:control_input_lanechange} appropriately one may place the non-zero characteristic roots to the left half complex plane and guarantee stability for the closed-loop system. 
In particular, using the control gains in Table~\ref{tab:cont_gains_lane} results in ${\lambda_k = -8}$\,s${^{-1}}$, ${k=1,\dots,6}$.

\begin{table}[!t]
	\caption{Control gains for the lane change maneuver}
	\label{tab:cont_gains_lane}
	\setlength{\tabcolsep}{1pt}
	\centering
	\begin{tabular}{llrl}
		\hline
		\multicolumn{4}{c}{\hspace{0.5ex}Below critical speed\hspace{0.5ex}}\\
		\multicolumn{4}{c}{\hspace{0.5ex}${v_{\mathrm G1} < v_{\mathrm{ G,crit}}}$\hspace{0.5ex}}                        
		\\ \hline
		\hspace*{0.5ex} 
		& $D_\vartheta$  \hspace{1eM} & $\mathbf{-1407.64}$ &\!Ns   \hspace{0.5ex} \\
		& $P_\vartheta$  \hspace{1eM} & $\mathbf{-7637.29}$ &\!N    \hspace{0.5ex} \\
		& $D_r$          \hspace{1eM} &           $2116.86$ &\!Ns/m \hspace{0.5ex} \\
		& $P_r$          \hspace{1eM} &          $11942.04$ &\!N/m  \hspace{0.5ex} \\
		& $P_\psi$       \hspace{1eM} &           $3382.02$ &\!N    \hspace{0.5ex} \\
		& $P_y$          \hspace{1eM} &           $4509.36$ &\!N/m  \hspace{0.5ex} \\ 
		\hline
	\end{tabular}
	~
	\begin{tabular}{llrl}
		\hline
		\multicolumn{4}{c}{\hspace{0.5ex}Above critical speed\hspace{0.5ex}}\\
		\multicolumn{4}{c}{\hspace{0.5ex}${v_{\mathrm G2} > v_{\mathrm{ G,crit}}}$\hspace{0.5ex}}                       \\ \hline
		\hspace*{0.5ex} 
		& $D_\vartheta$  \hspace{1ex} &   $\mathbf{105.36}$ &\!Ns   \hspace{0.5ex} \\
		& $P_\vartheta$  \hspace{1ex} &  $\mathbf{777.28}$ &\!N    \hspace{0.5ex} \\
		& $D_r$          \hspace{1ex} &            $99.52$ &\!Ns/m \hspace{0.5ex} \\
		& $P_r$          \hspace{1ex} &           $405.60$ &\!N/m  \hspace{0.5ex} \\
		& $P_\psi$       \hspace{1ex} &            $676.4$ &\!N    \hspace{0.5ex} \\
		& $P_y$          \hspace{1ex} &           $180.37$ &\!N/m  \hspace{0.5ex} \\ 
		\hline
	\end{tabular}
\end{table}

\begin{table}[!t]
	\caption{Control gains for the turning maneuver}
	\label{tab:cont_gains_turn}
	\setlength{\tabcolsep}{1pt}
	\centering
	\begin{tabular}{llrl}
		\hline
		\multicolumn{4}{c}{\hspace{0.5ex}Below critical speed\hspace{0.5ex}}\\
		\multicolumn{4}{c}{\hspace{0.5ex}${v_{\mathrm G1} < v_{\mathrm{ G,crit}}}$\hspace{0.5ex}}                       \\ \hline
		\hspace*{0.5ex} 
		& $D_\vartheta$  \hspace{1eM} & $\mathbf{-511.89}$ &\!Ns   \hspace{0.5ex} \\
		& $P_\vartheta$  \hspace{1eM} &       ${-2776.88}$ &\!N    \hspace{0.5ex} \\
		& $D_r$          \hspace{1eM} &           $882.53$ &\!Ns/m \hspace{0.5ex} \\
		& $P_r$          \hspace{1eM} &          $4881.84$ &\!N/m  \hspace{0.5ex} \\
		& $P_\psi$       \hspace{1eM} &           $536.67$ &\!N    \hspace{0.5ex} \\ \hline
	\end{tabular}
	~
	\begin{tabular}{llrl}
		\hline
		\multicolumn{4}{c}{\hspace{0.5ex}Above critical speed\hspace{0.5ex}}\\
		\multicolumn{4}{c}{\hspace{0.5ex}${v_{\mathrm G2} > v_{\mathrm{ G,crit}}}$\hspace{0.5ex}}                       \\ \hline
		\hspace*{0.5ex} 
		& $D_\vartheta$  \hspace{1ex} &  $\mathbf{118.88}$ &\!Ns   \hspace{0.5ex} \\
		& $P_\vartheta$  \hspace{1ex} &        ${-128.32}$ &\!N    \hspace{0.5ex} \\
		& $D_r$          \hspace{1ex} &            $41.49$ &\!Ns/m \hspace{0.5ex} \\
		& $P_r$          \hspace{1ex} &            $73.69$ &\!N/m  \hspace{0.5ex} \\
		& $P_\psi$       \hspace{1ex} &           $112.73$ &\!N    \hspace{0.5ex} \\ \hline
	\end{tabular}
	\vspace{-4mm}    
\end{table}

The performance  of the closed loop system is demonstrated by numerical simulations in the left column in Fig.~\ref{fig:manuvers}.
These were carried out using the full nonlinear equations~\eqref{eq:eqmot_unicycle} applying the control law~\eqref{eq:control_input_lanechange} with reference trajectory~\eqref{eq:ref_traj_lanechange_y} and control gains in Table~\ref{tab:cont_gains_lane}. 
The time evolution of tilt angle $\vartheta$, point mass position $r$, yaw angle $\psi$, and the control input $u$ are depicted in the top four panels, while the bottom panel shows the movement of the wheel center $ \rm G $ above the ${(x_0, y_0)}$-plane.
The desired lane change maneuver is followed by the unicycle, and the controller successfully  stabilizes the desired motion for both speeds.
The gains $P_\vartheta$ and $D_\vartheta$  are highlighted in Table~\ref{tab:cont_gains_lane} as they have opposite signs above and below the critical speed. 
Below the critical speed (${v_{\mathrm G1} < v_{\mathrm{ G,crit}}}$), the open loop dynamics is unstable, here the $P_\vartheta$ and $D_\vartheta$ gains correspond to positive stiffness and damping in the mechanical sense.
In contrast, above the critical speed (${v_{\mathrm G2} > v_{\mathrm{ G,crit}}}$) the open loop dynamics is stable and the $P_\vartheta$ and $D_\vartheta$ gains correspond to negative stiffness and damping. 
This illustrates that executing a lane change maneuver above the critical speed takes more effort, since the straight-rolling motion of the open-loop systems is more stable.
This can also be observed when comparing the 
required control input for the different speeds.
Still the required control input remains small (${|u|<10}$\,N)  for both speeds.

\subsubsection{Right turn}

For the 90$^\circ$ right turn maneuver, the unicycle eventually moves parallel to the $y_0$ axis and the state $y_{\rm G}$ becomes unreachable by the control input.
Thus, the state $y_{\rm G}$ may not be used to plan  motion in this case. 
Instead we use the yaw angle $\psi$ to construct the reference trajectory.
Considering this, the output vector 
\begin{equation}
	\label{eq:outputstates_90turn}
	\mathbf y = \mathbf C \mathbf x := \begin{bmatrix}
		\SIGMA_{1} 
		&\vartheta 
		&\SIGMAR 
		&r 
		&\psi 
	\end{bmatrix}^\mathsf T\!,
\end{equation}
with
\begin{equation}
	\mathbf C = \begin{bmatrix}
		1 & 0 & 0 & 0 & 0 & 0 & 0 & 0 & 0 & 0\\
		0 & 0 & 0 & 1 & 0 & 0 & 0 & 0 & 0 & 0\\
		0 & 0 & 0 & 0 & 1 & 0 & 0 & 0 & 0 & 0\\
		0 & 0 & 0 & 0 & 0 & 1 & 0 & 0 & 0 & 0\\
		0 & 0 & 0 & 0 & 0 & 0 & 1 & 0 & 0 & 0
	\end{bmatrix}
\end{equation}
is defined; cf.~\eqref{eq:output_lanechange} and \eqref{eq:output_lanechange_matrix}.
The output controllability matrix can be constructed as in \eqref{eq:contollability_matrix_output}; since it has maximal row rank, ${ \mathrm{rank}\,\mathbf M_{\rm oc} = 5 }$, the unicycle is output controllable.

Here the linear output feedback control law of the form \eqref{eq:control_input_lanechange} is applied 
with control gains ${\mathbf K = [D_\vartheta\ P_\vartheta\ D_r\ P_{r}\ P_{\psi}]}$ and 
\begin{equation}
	\label{eq:ref_traj_90turn}
	\arraycolsep=0pt
	\begin{array}{rcccccccccl}
		\mathbf y_{\rm des} = [
		&0
		&\hspace{1em}&0
		&\hspace{1em}&0
		&\hspace{1em}&0
		&\hspace{1em}&\psi_{\rm des}(t)
		&]^\mathsf{T},
	\end{array}  
\end{equation}
where the yaw angle
\begin{equation}
	\label{eq:ref_traj_turn}
	\psi_{\rm des}(t) = \begin{cases}
		\arraycolsep=0pt
		\begin{array}{lrl}
			0,               & 0\leq  &~t < 2,\\
			\frac{\pi}{4}\big({1-\cos\big(\tfrac{\pi}{5} (t-2)\big) }\big),\hspace{1em} & 2\leq  &~t<7,\\
			\frac{\pi}{2},               & 7\leq &~t < 10,\\
		\end{array}
	\end{cases}
\end{equation}
encodes the desired trajectory.
Such trajectory is depicted as a black dashed-dotted curve in Fig.~\ref{fig:manuvers}, (right column, third panel).
Choosing the control gains as in Table~\ref{tab:cont_gains_turn} for the speeds $v_{\mathrm G1}$ and $v_{\mathrm G2}$
the non-zero characteristic roots of the closed-loop system are placed at ${\lambda_k = -8}$\,s${^{-1}}$, ${k=1,\dots,5}$.
Since these are in the left half complex plane, the resulting closed-loop system is stable.

The simulation results are shown in the right side of Figure~\ref{fig:manuvers}.
These are carried out using the full nonlinear equations~\eqref{eq:eqmot_unicycle} applying the feedback law \eqref{eq:control_input_lanechange} with reference trajectory~\eqref{eq:ref_traj_90turn} and control gains in Table~\ref{tab:cont_gains_turn}.
The right turns are successfully executed for both speeds.
Comparing the control gains below and above the critical speed for the turning maneuver, only the derivative gain $D_\vartheta$ changes sign as highlighted in Table~\ref{tab:cont_gains_turn}.
When observing the time evolution the states and the control input in Figure~\ref{fig:manuvers}, one may notice some high-frequency content which is more pronounced for above the critical speed (orange curves).
These are due to the Coriolis force since the yaw rate is not zero while the point mass moves along the axle.
These effects are compensated by the controller while executing the maneuvers.

\section{Conclusion}\label{sec:conclusion}

In this study, modeling, analyses and control of an autonomous unicycle were considered.
The most compact form of the equations of motion were derived by means of the Appellian approach. 
The steady state motions of the rolling wheel were classified as cases of straight rolling, turning-rolling and spinning on the spot. 
The stability of these motions were determined by linear analysis and the critical angular velocities (above which the steady state motions are stable) were given in closed form. It was also shown that a turning-rolling steady state is always stable above a critical tilt angle which is independent of the system parameters.

Building upon the knowledge gained from the dynamics of the rolling wheel, the simplest possible control strategy is proposed for autonomous unicycles by means of actuating the position of a point mass normal to the wheel plane, in order to accomplish steering maneuvers such as lane changes and turns.  
The nonlinear equations of motions of the unicycle were given in closed form.
Apart from finding steady states akin to those of the rolling wheel, additional steady states were identified such as non-tilted turning and tilted spinning.
For the open-loop unicycle,  stability results were obtained for the straight rolling steady state and the stability of the turning-rolling steady states were determined semi-analytically, numerically.
All these stability results rely on linearization, the analysis of the conserved quantities is reserved for future research.%

Utilizing the equations linearized about the straight rolling steady state, two steering controllers of partial state feedback were proposed for the unicycle to carry out a lane change and a 90$^\circ$ turn. 
The resulting linear systems were shown to be output controllable, and when applying the designed controllers to the nonlinear system  the unicycle successfully completed the desired maneuvers  as confirmed by numerical simulations.

The developed modeling, analysis and control framework
may also allow more sophisticated motion planning as well as nonlinear control design which can take full advantage of the agility of unicycles.
Such techniques, which are left for future research, 
may allow one to achieve high level of maneuverability for autonomous unicycles and to provide steering assist for human-ridden unicycles.

The introduced unicycle model incorporates engineering assumptions, which do not necessarily hold in all circumstances. 
For example, sliding may occur violating the rolling assumption once the friction forces are not large enough.
This will be analyzed in a separate study.
Also, experiments are planned with the human-ridden and the autonomous unicycles to verify the theoretical findings.

\color{black}

\appendices

\section{Model derivation for the rolling wheel via the Appellian approach}\label{app:rollingdisc_model_deriv}

A vector resolved in frame ${\rm F}_{2}$ can be transformed to the frame ${\rm F}_{0}$ as 
\begin{equation} \label{eq:frame_transform}
	\square_{{\rm F}_{0}} = \mathbf T_{02} \square_{{\rm F}_{2}}\,,
\end{equation}
where the rotation matrix is 
\begin{equation} \label{eq:rot_matrix}
	\mathbf T_{02} = \begin{bmatrix}\cos{\psi} & - \sin{\psi} \cos{\vartheta} & \sin{\psi} \sin{\vartheta}
		\\
		\sin{\psi} & \cos{\psi} \cos{\vartheta} & - \cos{\psi} \sin{\vartheta} 
		\\
		0 & \sin{\vartheta} & \cos{\vartheta}\end{bmatrix}\,.
\end{equation}

The velocity of the wheel-ground contact point P can be expressed as
\begin{equation}\label{eq:kin_const_vect_eq}%
	\mathbf v_{\rm P} = \mathbf v_{\rm G} + \bm \omega \times \mathbf r_{\rm GP}\,, 
\end{equation}
where velocity of the center of gravity G can resolved in the ground-fixed frame ${\rm F}_0$ as
\begin{equation}
	\mathbf v_{\rm G} =  \dot {\mathbf r}_{\rm G} = 
	\begin{bmatrix} 
		\dot x_{\rm G} 
		& \dot y_{\rm G} 
		& \dot z_{\rm G} 
	\end{bmatrix}^{\mathsf T}_{{\rm F}_{0}}\,,%
	\label{eq:velocity}
\end{equation}
while the angular velocity of the wheel can be expressed 
in the moving frame ${\rm F}_2$ as
\begin{equation}
	\bm \omega = 
	\begin{bmatrix} 
		\dot\vartheta 
		& \dot\psi\sin\vartheta + \dot\varphi 
		& \dot\psi\cos\vartheta 
	\end{bmatrix}^{\mathsf T}_{{\rm F}_{2}}\,, \label{eq:omega}%
\end{equation}
using the tilt rate $\dot\vartheta$, the yaw rate $\dot\psi$, the pitch rate $\dot\varphi$ and the tilt angle $\vartheta$.
One may also express the position vector as
\begin{align} \label{eq:rPG}
	\mathbf r_{\rm GP} &= 
	\begin{bmatrix} 
		0 
		& 0 
		& -R 
	\end{bmatrix}^{\mathsf T}_{{\rm F}_{2}}\,.
\end{align}
Then transforming the cross product in \eqref{eq:kin_const_vect_eq} to the ground-fixed frame ${\rm F}_0$ with the help of \eqref{eq:rot_matrix},
the rolling condition~\eqref{eq:vP} results in the kinematic constraints \eqref{eq:kin_constr} and the geometric constraint \eqref{eq:geom_constr_eq_z}.

The definitions of pseudovelocities
are appropriate only if the generalized velocities 
($\dot x_{\mathrm G}$, $\dot y_{\mathrm G}$, $\dot \psi$, $\dot \vartheta$, $\dot \varphi$)
can unambiguously be expressed by means of the pseudovelocities 
($\SIGMA_1$, $\SIGMA_2$, $\SIGMA_3$)
and generalized coordinates 
($x_{\mathrm G}$, $y_{\mathrm G}$, $\psi$, $\vartheta$, $\varphi$).
To check this, we combine the kinematic constraints \eqref{eq:kin_constr} and the definitions of pseudovelocities \eqref{eq:sigmadef_disc} into one linear system of equations: 
\begin{equation}
	\arraycolsep=3.5pt
	\begin{bmatrix}
		0\\
		0\\
		\SIGMA_{1}\\    
		\SIGMA_{2}\\
		\SIGMA_{3}
	\end{bmatrix}
	\!=\!
	\underbrace{
		\begin{bmatrix}
			-1 & 0  & R \sin{\vartheta} \cos{\psi} & R \sin{\psi} \cos{\vartheta}   & R \cos{\psi} \\
			0  & -1 & R \sin{\psi} \sin{\vartheta} & - R \cos{\psi} \cos{\vartheta} & R \sin{\psi} \\
			0  & 0  & 0                            & 1                              & 0            \\
			0  & 0  & \sin\vartheta                & 0                              & 1            \\
			0  & 0  & \cos\vartheta                & 0                              & 0
		\end{bmatrix}
	}_{\displaystyle \mathbf C}
	\!%
	\begin{bmatrix}
		\dot x_{\rm G}\\
		\dot y_{\rm G}\\
		\dot\psi\\    
		\dot\vartheta\\
		\dot\varphi
	\end{bmatrix}\!.
\end{equation}
This linear system can be solved if $\mathbf C$ is invertible, that is, if its determinant is nonzero.
Since ${\det \mathbf C = \cos\vartheta}$, the matrix $\mathbf C$ is singular when the wheel is horizontal (${\vartheta = \pm \pi/2}$), which is excluded from this analysis.
Thus, the generalized velocities 
($\dot x_{\mathrm G}$, $\dot y_{\mathrm G}$, $\dot \psi$, $\dot \vartheta$, $\dot \varphi$) can be expressed with the pseudovelocities 
($\SIGMA_1$, $\SIGMA_2$, $\SIGMA_3$)
as
\begin{align}
	\label{eq:gen_velocs_disc}
	\begin{split}
		\dot x_{\rm G} &= \SIGMA_{1} R \sin{\psi} \cos{\vartheta} +\SIGMA_{2} R \cos{\psi}\,, 
		\\ 
		\dot y_{\rm G} &= - \SIGMA_{1} R \cos{\psi} \cos{\vartheta} + \SIGMA_{2} R \sin{\psi} \,, 
		\\
		\dot \psi &= \SIGMA_{3}\frac{1}{\cos{\vartheta}}\,, 
		\quad
		\dot \vartheta = \SIGMA_{1}\,, 
		\quad
		\dot \varphi = \SIGMA_{2} - \SIGMA_{3} \tan{\vartheta}\,. 
	\end{split}
\end{align}
Also, the velocity of the center of gravity G \eqref{eq:velocity} and the angular velocity \eqref{eq:omega} can be expressed in the terms of pseudovelocities ($\SIGMA_1$, $\SIGMA_2$, $\SIGMA_3$) using \eqref{eq:gen_velocs_disc}; these vectors have the most compact form when expressed in frame ${\rm F}_{2}$:
\begin{equation}
	\begin{split}
		\mathbf v_{\rm G} &= \begin{bmatrix} 
			\dot \psi R \sin{\vartheta\,\!}  + \dot \varphi R
			& - \dot \vartheta  R 
			& 0
		\end{bmatrix}^{\mathsf T}_{\rm F_2} \equiv \begin{bmatrix} 
			\omega_{2}R 
			& - \omega_{1} R
			& 0
		\end{bmatrix}^{\mathsf T}_{\rm F_2},
		\\
		\bm \omega &= \begin{bmatrix} 
			\dot\vartheta 
			& \,\dot\psi\sin\vartheta + \dot\varphi 
			& \dot\psi\cos\vartheta 
		\end{bmatrix}^{\mathsf T}_{{\rm F}_{2}} \equiv \begin{bmatrix} 
			\omega_{1}
			& \omega_{2}
			& \omega_{3}
		\end{bmatrix}^{\mathsf T}_{\rm F_2}. 
	\end{split}
\end{equation}

The acceleration energy of a rigid body is defined as:
\begin{equation}\label{eq:accel_energy_wheel}
	S = \frac{1}{2} m \mathbf a_{\rm G}^2 + \frac{1}{2}\bm\alpha \cdot \mathbf J_{\rm G} \bm\alpha + \bm\alpha \cdot ( \bm\omega \times \mathbf J_{\rm G} \bm\omega )\,,
\end{equation}
where ${\mathbf a_{\rm G} = \dot {\mathbf v}_{\rm G}}$ is the acceleration of the center of gravity, ${\bm\alpha=\dot{\bm\omega}}$ is the angular acceleration and $\mathbf J_{\rm G}$ is the mass moment of inertia with respect to the center of gravity.
All these vectors must be expressed based on the generalized coordinates ($x_{\mathrm G}$, $y_{\mathrm G}$, $\psi$, $\vartheta$, $\varphi$),
pseudovelocities ($\SIGMA_1$, $\SIGMA_2$, $\SIGMA_3$)
and pseudoaccelerations ($\dot \SIGMA_1$, $\dot \SIGMA_2$, $\dot \SIGMA_3$).
Also, the vectors have the most compact form when expressed in frame ${\rm F}_{2}$:
\begin{align}
	\mathbf a_{\rm G} &= 
	\begin{bmatrix}
		R (\dot \SIGMA_{2} + \SIGMA_{1} \SIGMA_{3}) \\
		- R (\dot \SIGMA_{1} - \SIGMA_{2} \SIGMA_{3}) \\
		- R (\SIGMA_{1}^{2} + \SIGMA_{2} \SIGMA_{3} \tan{\vartheta})
	\end{bmatrix}_{{\rm F}_{2}}\!\!,%
	\ \ 
	\mathbf J_{\rm G} = \frac{mR^2}{4}
	\begin{bmatrix}
		1 & 0 & 0\\
		0 & 2 & 0\\
		0 & 0 & 1
	\end{bmatrix}_{{\rm F}_{2}}\!\!,%
	\nonumber\\
	&\qquad\qquad\bm\alpha = 
	\begin{bmatrix}
		\dot \SIGMA_{1} - \SIGMA_{2} \SIGMA_{3} + \SIGMA_{3}^{2} \tan{\vartheta}\\
		\dot \SIGMA_{2}\\
		\dot \SIGMA_{3} + \SIGMA_{1} \SIGMA_{2} - \SIGMA_{1} \SIGMA_{3} \tan{\vartheta}
	\end{bmatrix}_{{\rm F}_{2}}\!\!.
	\label{eq:mech_properties_disc}%
\end{align}
The calculation of the accelerations is not trivial in the moving frame $\mathrm F_2$; still, this gives the simplest possible algebraic form.

These lead to the acceleration energy of the rolling disc:
\begin{equation}
	\label{eq:acc_energy_disc}
	\begin{split}
		S =&\, \frac{m R^2}{8} \big(
		5 \dot \SIGMA_{1}^{2} + 6 \dot \SIGMA_{2}^{2} + \dot \SIGMA_{3}^{2} 
		+ (2 \SIGMA_{3}^{2} \tan{\vartheta} - 12 \SIGMA_{2} \SIGMA_{3})\dot \SIGMA_{1}
		\\&
		+ 8 \SIGMA_{1} \SIGMA_{3} \dot \SIGMA_{2}
		+ (4 \SIGMA_{1} \SIGMA_{2} - 2 \SIGMA_{1} \SIGMA_{3} \tan{\vartheta}) \dot \SIGMA_{3} 
		\big)
		+ \dots\,, 
	\end{split}
\end{equation}
where the dots ($\dots$) represent further terms that are independent of the pseudoaccelerations 
($\dot \SIGMA_1$, $\dot \SIGMA_2$, $\dot \SIGMA_3$)
so they can be neglected.

Appell's equations are formulated as
\begin{equation}
	\dfrac{\partial S}{\partial \dot\SIGMA_{\!j}} = \Pi_j, \quad j=1,\dots,3\,, \label{eq:appell}
\end{equation}
where $\Pi_j$ is the pseudoforce corresponding to the pseudoacceleration $\dot\SIGMA_j$.
The pseudoforces can be calculated from the virtual power of the active forces:
\begin{equation}
	\delta P = \mathbf G \cdot \delta \mathbf v_{\rm G} = \sum_{j=1}^3 \Pi_j\,\delta \SIGMA_{\!j}\,.
\end{equation}
Here ${\mathbf G = [0\ \ 0\ \ -mg]_{{\rm F}_{0}}^\mathsf T }$ represents the gravitational force, the only active force in our model, while ${\delta \mathbf v_{\rm G} =  [\delta \dot x_{\rm G}\ \,\delta \dot y_{\rm G}\ \,\delta \dot z_{\rm G}]_{{\rm F}_{0}}^\mathsf T =  [\ \cdot\ \ \cdot \ \ -R \sin{\vartheta}\, \delta \SIGMA_{1}]_{{\rm F}_{0}}^\mathsf T}$ represents the virtual velocity; cf.~\eqref{eq:velocity}. These yield the pseudoforces:
\begin{align}
	\Pi_1 &=m g R\sin{\vartheta}
	\,,\quad 
	\Pi_2 = 0  \label{eq:pseudo_forces_disc}
	\,,\quad 
	\Pi_3 = 0\,.
\end{align}

Based on the acceleration energy \eqref{eq:acc_energy_disc}, Appell's formula \eqref{eq:appell} and the pseudoforces \eqref{eq:pseudo_forces_disc}, the pseudoaccelerations can be expressed as
\begin{align}\label{eq:eq:dsigma_eqs}
	\begin{split}
		\dot{\SIGMA}_1 &= \frac{6}{5} \SIGMA_2 \SIGMA_3 - \frac{1}{5} \SIGMA_{3}^{2} \tan \vartheta + \frac{4 g}{5 R} \sin\vartheta \,, %
		\\
		\dot{\SIGMA}_2 &= - \frac{2}{3} \SIGMA_1 \SIGMA_3\,,
		\\
		\hspace{0.1ex}\dot{\SIGMA}_3 &= - 2 \SIGMA_1 \SIGMA_2 + \SIGMA_1 \SIGMA_3 \tan\vartheta\,\!,          
	\end{split}
\end{align}
These equations together with the generalized velocities \eqref{eq:gen_velocs_disc} form the system of eight first order ordinary differential equations, which constitute the equations of the motion of the rolling wheel; cf.~\eqref{eq:eq_mot_disc}.

\section{Stability of the rolling wheel}\label{app:disc_stability}

To prove the stability of the steady states of the rolling wheel, we follow \cite{NeiFuf1972}. 
First, time is eliminated from the second and third equations of the essential motion \eqref{eq:eq_mot_disc} by expressing
\begin{equation} \label{eq:domega_dtheta}
	{\omega}_{2}^{\prime} = - {2} \omega_{3} / {3}
	\,,
	\quad
	{\omega}_{3}^{\prime} = - 2 \omega_{2} + \omega_{3} \tan{\vartheta\,\!}
	\,,
\end{equation}
where ${ {\omega}_{2}^{\prime} = \frac{\mathrm d{\omega}_{2}}{\mathrm d \vartheta } }$, ${ {\omega}_{3}^{\prime} = \frac{\mathrm d{\omega}_{3}}{\mathrm d \vartheta }}$ and ${\dot\vartheta = \omega_{1}}$.
Since the wheel is a conservative mechanical system, the total energy 
${%
	E = T+U  %
}$
is constant,
${
	E(t) \equiv E_{*}\,,
}$
where 
${%
	T = m R^2 \big(5  \omega_{1}^{2} + 6  \omega_{2}^{2} +  \omega_{3}^{2} \big)/{8} %
}$ 
and 
${%
	U = m g R  \cos{\vartheta\,\!} %
}$
are the kinetic and potential energies of the wheel, respectively.
Rearranging the total energy gives
\begin{equation} 
	\omega_{1}^2 \equiv \dot\vartheta^2 =  \underbrace{
		\frac{1}{5 m R^2}\big({8 E_{*} - m R (6 \omega_{2}^{2} R \!-\!\omega_{3}^{2} R \!-\!8 g \cos{\vartheta\,\!}})\big)
	}_{
		\displaystyle
		= f(\vartheta;\,\vartheta_{\mathrm{ic}},\omega_{2,\mathrm{ic}},\omega_{3,\mathrm{ic}},\,E_{*} )
	}.
\end{equation}

The condition for stability is that ${f}$ has a strict maximum at ${\vartheta = \vartheta_{*}}$.
Expressing
${ 
	{\mathrm{d} f}/{\mathrm{d} \vartheta} = 0 
}$ 
gives back formula \eqref{eq:steadystaterelation_genveloc_disc} representing the steady state surface in Figure~\ref{fig:levels_and_stability_disc}.
The necessary and sufficient condition for steady states to be stable can be obtained with
${
	{\mathrm{d}^2 f}/{\mathrm{d} \vartheta^2} < 0
}$%
, where \eqref{eq:domega_dtheta} is substituted.
This leads to \eqref{eq:wheelNeccAndSuffCond} which is obtained from analyzing the linearized systems.

\color{black}

\section{Model derivation for the unicycle via the Appellian approach}\label{app:unicycle_model_deriv}

With the added point mass the definition \eqref{eq:pseudoveloc_defs_unicycle} of pseudovelocities is appropriate because the generalized velocities 
($\dot x_{\mathrm G}$, $\dot y_{\mathrm G}$, $\dot \psi$, $\dot \vartheta$, $\dot \varphi$, $\dot r$)
can unambiguously be expressed by means of the pseudovelocities ($\SIGMA_1$, $\SIGMA_2$, $\SIGMA_3$, $\SIGMAR$)
and the generalized coordinates ($x_{\mathrm G}$, $y_{\mathrm G}$, $\psi$, $\vartheta$, $\varphi$, $r$)
when \eqref{eq:gen_velocs_disc} is extended with ${\dot r = \SIGMAR + \omega_{1}R}$.

The acceleration energy $S$ of the unicycle consisting of the rigid wheel and the point mass is defined as:
\begin{equation}
	S = \frac{1}{2} m \mathbf a_{\rm G}^2 
	+ \frac{1}{2}\bm\alpha\cdot \mathbf J_{\rm G} \bm\alpha 
	+ \bm\alpha\cdot ( \bm\omega \times  \mathbf J_{\rm G} \bm\omega ) +  \frac{1}{2} m_0 \mathbf a_{\rm A}^2 \,,
\end{equation}
where the first three terms are identical to those in \eqref{eq:accel_energy_wheel} and \eqref{eq:mech_properties_disc}. 
The velocity \(	\mathbf v_{\!\rm A}\) of the point mass \(m_0\) can be expressed as 
\begin{align}
	\mathbf v_{\!\rm A} = \begin{bmatrix}
		\dot \psi (R \sin{\vartheta\,\!} - r \cos{\vartheta\,\!}) + \dot \varphi  R  
		\\
		\dot r - \dot \vartheta R  
		\\
		\dot \vartheta r
	\end{bmatrix}_{{\rm F}_{2}}
	\equiv
	\begin{bmatrix}
		\omega_{2} R  - \omega_{3} r
		\\
		\sigma
		\\
		\omega_{1} r
	\end{bmatrix}_{{\rm F}_{2}} \!,
\end{align}	
while the acceleration \(\mathbf a_{\rm A}\) becomes
\begin{align}
	\!\mathbf a_{\rm A} = \begin{bmatrix}
		\dot \omega_{2}  R  - \dot \omega_{3} r  +  \omega_{1} \omega_{3}  \left(r \tan{\vartheta\,\!} - R\right)- 2 \omega_{3}\sigma 
		\\
		\dot \sigma - \omega_{1}^{2} r - \omega_{3}^{2} r  + \omega_{2} \omega_{3} R \ \ \ 
		\\
		\dot \omega_{1} r + \omega_{1}^{2} R  + 2 \omega_{1}  \sigma  + \omega_{3}^{2} r \tan{\vartheta\,\!} - \omega_{2} \omega_{3} R  \tan{\vartheta\,\!} 
	\end{bmatrix}_{{\rm F}_{2}} \!\!\!.
\end{align}
These lead to
\begin{equation}
	\label{eq:acc_energy_unicycle}
	\begin{split}
		S =& 
		+ \frac{1}{8}\big({5 m R^2} + 4{m_{0} r^{2}}\big) \dot \omega_{1}^{2} 
		+ \frac{1}{4}\big({3 m } + 2{m_{0}  }\big) R^2 \dot \omega_{2}^{2} 
		\\&
		+ \frac{1}{8}\big({m R^2} + {4 m_{0} r^{2}}\big) \dot \omega_{3}^{2}
		+ \frac{m_{0}}{2} \dot \sigma^{2}
		- m_{0} R  r \dot \omega_{2} \dot \omega_{3} 
		\\&
		+ \frac{1}{4} \big( 
		4 \omega_{1}^{2} m_{0} R   r 
		+ 8  \omega_{1} \sigma  m_{0} r 
		+ \omega_{3}^{2} ({m R^2} + 4 m_{0} r^{2} ) \tan{\vartheta\,\!}
		\\&\qquad
		- \omega_{2} \omega_{3}  ( {6 m R^2} + 4 m_{0} R  r \tan{\vartheta\,\!}) 
		\big) \dot \omega_{1}
		\\&
		+ \big( 
		\omega_{1} \omega_{3} (m R^2 - m_{0} R^2  + m_{0} R  r \tan{\vartheta\,\!})  
		- 2  \omega_{3} \sigma  m_{0} R 
		\big) \dot \omega_{2} 
		\\&
		+ \frac{1}{4}  \big( 
		{2 \omega_{1} \omega_{2} m R^2 }
		+ \omega_{1} \omega_{3} (
		4 m_{0} R  r 
		- {m R^2 \tan{\vartheta\,\!}} 
		\\&\qquad
		- 4  m_{0} r^{2} \tan{\vartheta\,\!}) 
		+ 8  \omega_{3} \sigma m_{0}  r 
		\big) \dot \omega_{3} 
		\\&
		+ \big( \omega_{2} \omega_{3} m_{0} R  
		- \omega_{1}^{2}  m_{0} r - \omega_{3}^{2} m_{0}  r\big) \dot \sigma
		+ \dots\,.
	\end{split}
\end{equation}
The pseudoforces $\Pi_j$ are expressed based on the virtual power:
\begin{equation}
	\delta P = \mathbf G \cdot \delta \mathbf v_{\rm G} + \mathbf G_{\!\rm A} \cdot \delta \mathbf v_{\!\rm A} +
	\mathbf F \cdot \delta\mathbf v_{\!\rm A} - \mathbf F \cdot \delta\mathbf v_{\rm G}
	= \sum_{j=1}^4 \Pi_j\, \delta \SIGMA_{\!j}\,,
	\label{eq:virtualpower_unicycle}
\end{equation}
where
${\mathbf G = [0\ \ 0\ \ -mg]_{{\rm F}_{0}}^\mathsf T }$ and ${\mathbf G_{\rm A} = [0\ \,0\ -\!m_0g]_{{\rm F}_{0}}^\mathsf T}$ are the gravitational forces acting on the wheel and the point mass, respectively,
\({
	\delta \mathbf v_{\!\rm A} = [
	R\hspace{0.3mm} \delta 	\omega_{2} -  r\hspace{0.3mm}  \delta \omega_{3}
	\ \ 
	\delta\sigma
	\ \ 
	r\hspace{0.3mm}  \delta\omega_{1} 
	]_{{\rm F}_{2}}^{\mathsf T} ,
}\)
$\mathbf F$ is the control force \eqref{eq:Fr} and we used the notation ${\SIGMA_4 = \sigma}$.
This yields the following pseudoforces:
\begin{equation}
	\label{eq:pseudo_forces_unicycle}
	\begin{split}
		\Pi_1 &= ( m + m_0 ) g R \sin{\vartheta} -  m_0 g r \cos{\vartheta}\,,\\
		\Pi_2 &= 0\,,\quad
		\Pi_3 = 0\,,\quad
		\Pi_4 = u -  m_0 g \sin{\vartheta}\,. 
	\end{split}
\end{equation}

In case of ${m_0> 0}$, the pseudoaccelerations ($\dot \SIGMA_1$, $\dot \SIGMA_2$, $\dot \SIGMA_3$, $\dot \SIGMAR$)
can be obtained from the Appellian equations of the form~\eqref{eq:appell}, while the generalized velocities ($\dot x_{\mathrm G}$, $\dot y_{\mathrm G}$, $\dot \psi$, $\dot \vartheta$, $\dot \varphi$, $\dot r$)
can also be expressed from the definition of the pseudovelocities~\eqref{eq:pseudoveloc_defs_unicycle} and the kinematic constraints~\eqref{eq:kin_constr} for $ \vartheta\neq \pm \pi/2 $. 
These result in the equations of motion of the unicycle \eqref{eq:eqmot_unicycle}.

\vspace{-0.5ex}
\section{Elements of matrix $\mathbf A$ in \eqref{eq:Amat_rollturn_unicycle}}\label{app:Amatgeneral_elements_unicycle}

Note that in the following formulas the steady state tilt angle $\vartheta_{*}$, yaw rate $\dot{\psi}_{*}$, pitch rate $\dot{\varphi}_{*}$ and point mass position $r_{*}$ are not independent of each other but they must satisfy the relations in~\eqref{eq:steadystaterelation_genveloc_unicycle}:
\begingroup
\allowdisplaybreaks
\begin{subequations}
	\begin{align*}
		A_{12} &= \frac{\dot\psi_{*} \left(6 m R^2 \cos{\vartheta_{*} } + 4 m_{0} R  r_{*}  \sin{\vartheta_{*} }\right)}{5 m R^2 + 4 m_{0} r_{*}^{2}}
		\\
		A_{13} &= \frac{1}{5 m R^2 + 4 m_{0} r_{*}^{2}}\big(
		\dot\varphi_{*} R (6 m R + 4  m_{0} r_{*} \tan{\vartheta_{*}  }) 
		\\&\quad
		+ \dot\psi_{*} (4 m_{0} R  r_{*} \tan{\vartheta_{*} } + (4 m R^2 - 8 m_{0} r_{*}^{2}) ) \sin{\vartheta_{*} }
		\big)
		\\
		A_{14} &= \frac{1}{\left(5 m R^2 + 4 m_{0} r_{*}^{2}\right) \cos{\vartheta_{*} }}\big(
		4 \dot\psi_{*} \dot\varphi_{*} m_{0} R r_{*} 
		\\&\quad
		+ \dot\psi_{*}^{2} (4 m_{0} R  r_{*}  \tan{\vartheta_{*} } 
		- ( m R^2 + 4 m_{0} r_{*}^{2}) ) \cos{\vartheta_{*} } 
		\\&\quad
		+ 4 m_{0} g r_{*} \sin{\vartheta_{*} } \cos{\vartheta_{*} }
		+ 4 m g R \cos^{2}{\vartheta_{*} } 
		\big)
		\\
		A_{16} &= \frac{m_{0}}{\left(5 m R^2 + 4 m_{0} r_{*}^{2}\right)^{2}}\big(
		(16  m_{0} g r_{*}^{2} - 20 m  g R^{2} ) \cos{\vartheta_{*} }	
		\\&\quad
		+ \dot\psi_{*} \dot\varphi_{*} ( (20   m  R^{3} - 16  m_{0}  R r_{*}^{2}) \sin{\vartheta_{*} }
		- 48  m  R^{2} r_{*} \cos{\vartheta_{*} } )
		\\&\quad		
		- 32 m  g R r_{*} \sin{\vartheta_{*} }
		+ 4 \dot\psi_{*}^{2} (
		(5 m R^3  - 4 m_{0} R r_{*}^{2}) \sin^{2}{\vartheta_{*}  }
		\\&\qquad
		- 20  m R^2 r_{*} \sin{\vartheta_{*} } \cos{\vartheta_{*} } 
		)  
		\big)
		\\
		A_{21} &= \frac{1}{3 m R^2 + 2 m_{0} R^2  + 12 m_{0} r_{*}^{2}}\big(
		- 4 R \dot\varphi_{*} m_{0} r_{*} 
		\\&\quad
		+ 2 \dot\psi_{*} (( m_{0} R^2-  m R^2 - 4 m_{0} r_{*}^{2}) \cos{\vartheta_{*} } - 2 m_{0} R  r_{*}  \sin{\vartheta_{*} })
		\big)
		\\
		A_{25} &= \frac{4 R \dot\psi_{*} m_{0} \cos{\vartheta_{*} }}{3 m R^2 + 2 m_{0} R^2  + 12  m_{0} r_{*}^{2}}
		\\
		A_{31} &= \frac{1}{3 m R^2 + 2 m_{0} R^2  + 12 m_{0} r_{*}^{2}}\big(
		- \dot\varphi_{*} ( 6  m R^{2}  + 4 m_{0} R^{2} )
		\\&\quad
		+ \dot\psi_{*} ((12 m_{0} r_{*}^{2} - 3 m R^2 - 2 m_{0} R^2 ) \sin{\vartheta_{*} } 
		\\&\qquad
		- 20  m_{0} R  r_{*} \cos{\vartheta_{*} })
		\big)
		\\
		A_{35} &= - \frac{24 \dot\psi_{*} m_{0} r_{*} \cos{\vartheta_{*} }}{3 m R^2 + 2 m_{0}  R^2  + 12 m_{0} r_{*}^{2}}
		\\
		A_{52} &= - \dot\psi_{*} R \cos{\vartheta_{*} } 
		\,,\hspace{1ex}
		A_{53} = \dot\psi_{*} ( 2 r_{*}  \cos{\vartheta_{*} } - R \sin{\vartheta_{*} } ) - \dot\varphi_{*}  R 
		\\
		A_{54} &= - g \cos{\vartheta_{*} } 
		\,,\quad
		A_{56} = \dot\psi_{*}^{2} \cos^{2}{\vartheta_{*} } 
		\,,\quad
		A_{61} = R 
		\\   %
		A_{73} &= \frac{1}{\cos{\vartheta_{*}}}  
		\,,\quad %
		A_{74} = \dot{\psi}_{*} \tan{\vartheta_{*}}\,,\quad%
		A_{83} = - \tan{\vartheta_{*}}\,,\\
		A_{84} &= - \frac{\dot{\psi}_{*}}{\cos{\vartheta_{*}}}\,,\quad%
		A_{91} = R \sin{\psi_{*}} \cos{\vartheta_{*}}\,,\quad%
		A_{92} = R \cos{\psi_{*}}\,,\\%[1ex]
		A_{97} &= - R \left(\dot{\psi}_{*} \sin{\vartheta_{*}} + \dot{\varphi}_{*}\right) \sin{\psi_{*}}\,,\quad%
		A_{101} = - R \cos{\psi_{*}} \cos{\vartheta_{*}}\,,\\
		A_{102} &= R \sin{\psi_{*}}\,, \quad
		A_{107} = R \left(\dot{\psi}_{*} \sin{\vartheta_{*}} + \dot{\varphi}_{*}\right) \cos{\psi_{*}}\,.\ %
	\end{align*}
\end{subequations}
\endgroup

\vspace{-12mm}
\begin{IEEEbiography}[{\includegraphics[width=1in,height=1.25in,clip,keepaspectratio]{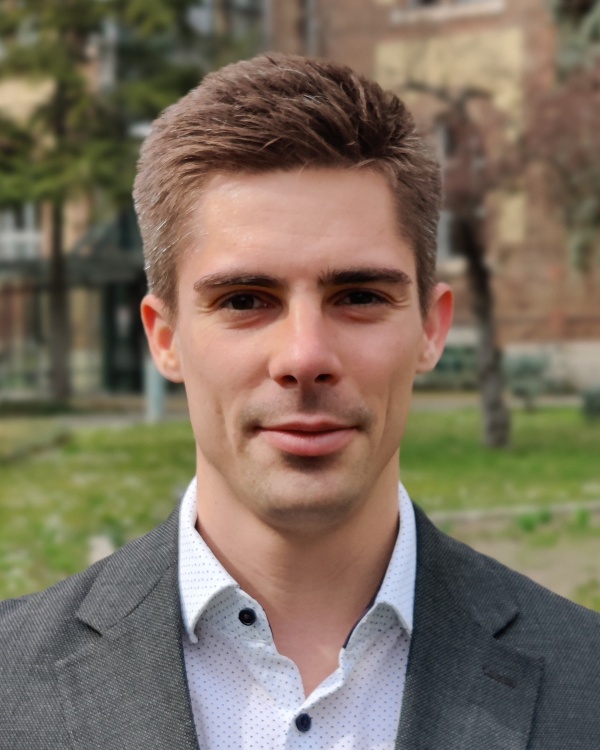}}]
	{M\'at\'e B. Vizi} received the BSc~degree in Mechatronic Engineering and the MSc~degree in Mechanical Engineering Modelling
	from the Budapest University of Technology and Economics, Hungary, in 2016 and 2018, respectively. He received the PhD~degree in the same institution in 2024.
	Currently he has a postdoctoral position at the University of Michigan. His research interests include nonlinear dynamics, control and time delay systems.
\end{IEEEbiography}

%\vspace{-14mm}
\begin{IEEEbiography}[{\includegraphics[width=1in,height=1.25in,clip,keepaspectratio]{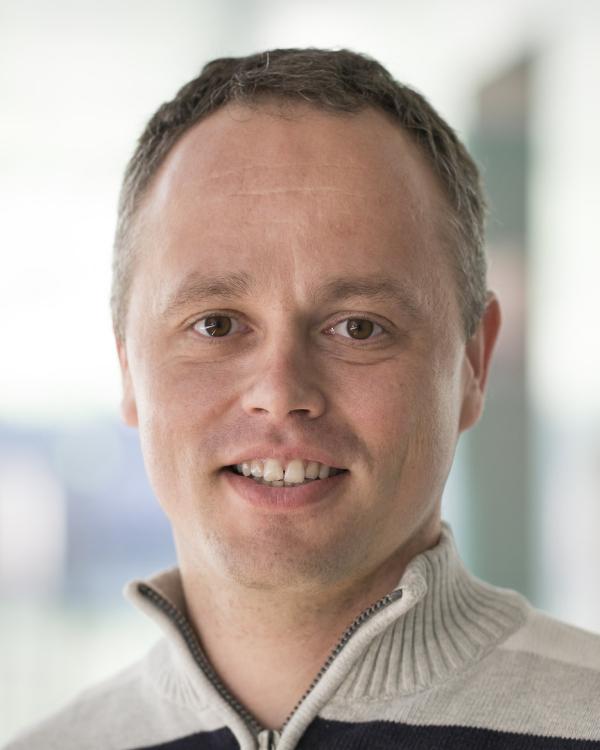}}]{G{\'{a}}bor Orosz} received the MSc degree in Engineering Physics from the Budapest University of Technology, Hungary, in 2002 and the PhD degree in Engineering Mathematics from the University of Bristol, UK, in 2006. 
	He held postdoctoral positions at the University of Exeter, UK, and at the University of California, Santa Barbara. 
	In 2010, he joined the University of Michigan, Ann Arbor where he is currently a Professor in Mechanical Engineering and in Civil and Environmental Engineering. 
	From 2017 to 2018 he was a Visiting Professor in Control and Dynamical Systems at the California Institute of Technology. 
	In 2022 he was a Distinguished Guest Researcher in Applied Mechanics at the Budapest University of Technology and from 2023 to 2024 he was a Fulbright Scholar at the same institution. 
	His research interests include nonlinear dynamics and control, time delay systems, machine learning, and data-driven systems with applications to connected and automated vehicles, traffic flow, and biological networks.
\end{IEEEbiography}

%\vspace{-12mm}
\begin{IEEEbiography}[{\includegraphics[width=1in,height=1.25in,clip,keepaspectratio]{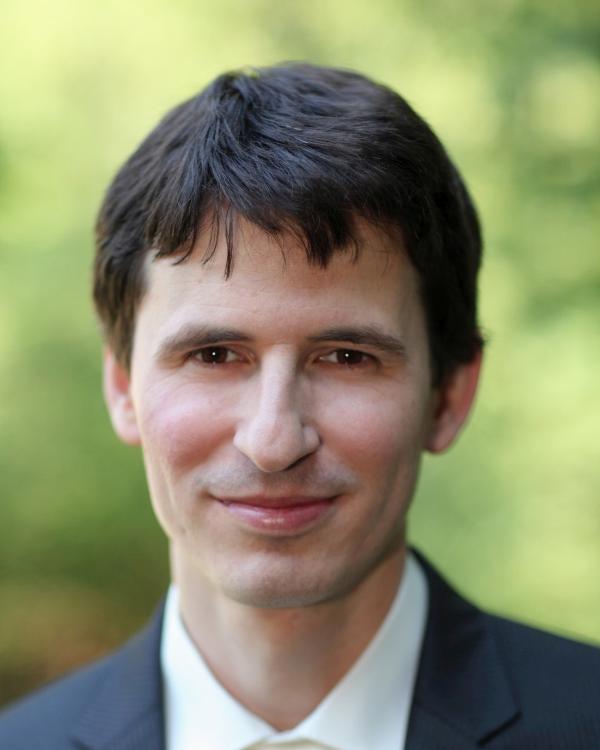}}]{D\'enes Tak\'acs} received his MSc and PhD in Mechanical Engineering from the Budapest University of Technology and Economics in 2005 and 2011, respectively. Between 2011 and 2018, he worked in the MTA-BME Research Group on Dynamics of Machines and Vehicles in Budapest, Hungary. Since 2018, he has been an Associate Professor at Budapest University of Technology and Economics, Budapest, Hungary. His research interests include tire and vehicle dynamics, nonlinear dynamics and time delay systems.
\end{IEEEbiography}

%\vspace{-12mm}
\begin{IEEEbiography}[{\includegraphics[width=1in,height=1.25in,clip,keepaspectratio]{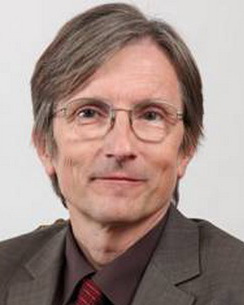}}]
	{G\'abor St\'ep\'an} received the MSc~and PhD~degrees in mechanical engineering from Budapest University of Technolgy and Economics, Hungary, in 1978 and 1982, respectively, and the DSc~degree from the Hungarian Academy of Sciences, Budapest, Hungary, in 1994. 
	He was a Visiting Researcher in the Mechanical Engineering Department of the University of Newcastle upon Tyne, UK, during 1988--1989, the Laboratory of Applied Mathematics and Physics of the Technical University of Denmark in 1991, and the Faculty of Mechanical Engineering of the Delft University of Technology during 1992--1993. 
	He was a Fulbright Visiting Professor at the Mechanical Engineering Department of the California Institute of Technology during 1994--1995, and a Visiting Professor at the Department of Engineering Mathematics of Bristol University in 1996. 
	He is currently a Professor of Applied Mechanics at the Budapest University of Technology and Economics. He is a fellow of CIRP and SIAM, received the the Delay Systems Lifetime Achievements Award of IFAC, the Caughey Dynamics Award and the Lyapunov Award of ASME.
	He is a member of the Hungarian Academy of Sciences and the Academy of Europe.
	His research interests include nonlinear vibrations in delayed dynamical systems, and applications in mechanical engineering and biomechanics such as wheel dynamics (rolling, braking, shimmy), robotic force control, machine tool vibrations, human balancing, and traffic dynamics.
	
\end{IEEEbiography}

\vfill

\end{document}